\documentclass[12pt]{article}
\usepackage{mathrsfs}
\usepackage{}
\usepackage{eurosym}
\usepackage{amssymb,amsmath,epsfig}
\usepackage{color}
\usepackage{amsmath,amsfonts,amssymb}
\usepackage{geometry}
\begin{document}

\title{\bf Strong Gravitational Lensing for Photon Coupled to Weyl Tensor in Kiselev Black Hole}
\author{ G. Abbas$^{a}$ \thanks{abbasg91@yahoo.com}, Asif Mahmood$^{a}$
\thanks{mahmoodasif486@gmail.com} and M. Zubair$^{b}$\thanks{drmzubair@cuilahore.edu.pk}\\
$^{a}$Departmentof Mathematics, The Islamia University of\\
Bahwalpur, Pakistan\\
$^{b}$Department of Mathematics, COMSATS University,\\ Islamabad, Lahore Campus, Pakistan}
\date{}

\maketitle
\begin{abstract}
The ambition of the present work is to highlight the phenomena of strong gravitational lensing and deflection angle for the photons coupling with Weyl tensor in a Kiselev black hole. Here, we have extended the prior work of Chen and Jing \cite{1} for Schwarzschild black hole to Kiselev black hole.
For this purpose, the equation of motion for the photons coupled to Weyl tensor, null geodesic and equation of photon sphere in a
Kiselev black hole spacetime have been formulated. It is found that the equation of motion of the photons depends not only on the coupling between photon and Weyl tensor, but also on the polarization direction of the photons. There is a critical value of the coupling parameter $\alpha$ for existence of the marginally circular
photon orbit outside the event horizon, which depends on the parameters of black hole
and the polarization direction of photons. Further, the polarization directions of coupled photon and the coupling parameter $\alpha$, both modify the features of the photon sphere, the angle of deflection and the functions $(\bar{a}$ and $\bar{b})$ for the strong gravitational lensing in Kiselev black hole spacetime.
In addition to this, the observable gravitational lensing quantities and the shadows of the Kiselev black hole spacetime are presented in detail.\\
\textbf{PACS numbers}: 95.30.Sf, 04.20.Dw, 04.70.Bw, 98.62.S\\
\textbf{Keywords}: Relativity and Gravitation; Gravitational lensing; Classical black holes; Deflection angle.
\end{abstract}

\section{Introduction}
 The interaction between gravitational and electromagnetic forces are of the vital importance in
material science due to the fact that in nature both gravity and electromagnetic force are two
major types of the basic forces. In the Lagrangian of Einstein-Maxwell field theory, the second
order term in the Maxwell tensor is associated directly to the gravitational and electromagnetic forces. Moreover, the interaction between the Riemann curvature tensor of the spacetime and Maxwell field
are not coupled in electromagnetic theory. However, such type of the coupling in quantum electrodynamics (QED) discovered by Drummond $et~al$.\cite{2} can be observed clearly by the effective action of the photons, created through one-loop vacuum polarization over a background curved spacetime.
The coupling between the Riemann curvature tensor and the electromagnetic field is
just a quantum phenomenon. In the effective field theory, all of the coupling constant terms are
tiny. Therefore, their values should be of the order of the square of the Compton wave-length of the
electron $\lambda_{e}$. Some authors \cite{3,4} have explored the certain fascinating effects over the electromagnetic variances by reconsidering the structure of Drummond \cite{2} along temporary coupled constant functions. Ni \cite{5} arranged a standard electromagnetic model, by considering that the coupling between
electromagnetic field and curvature tensor should be emerged reasonably in the region near the classical
supermassive compact objects at the center of galaxies due to their strong gravity and high mass density, Ni's
model has been investigated widely in astrophysics \cite{6,7} and black hole physics \cite{8,9}.
It has been shown by Ritz and Ward \cite{10} that, in the electrodynamics with a
Weyl correction, the universal relation with the U(1) central charge is changed
for the holographic conductivity in the background of anti-de Sitter spacetime.
Similarly, the critical temperature and the order of the phase transition are
modified in the formation of the holographic superconductor because of the
presence of such coupling terms \cite{11}-\cite{15}. Moreover, with these couplings it has been shown in \cite{16}, that the dynamical
evolution and Hawking radiation of the electromagnetic field in the black hole
spacetime depend on the coupling parameter and the parity of the field.
A nice way to give an explanation for the power-law expansion within the earliest Universe and the electromagnetic forces discovered at wide scale inside the galaxies clusters. Such sorts of observations suggest that the coupled terms for both gravitational and electromagnetic fields change the equation of motion. The time delay within the arriving of both electromagnetic and gravitational waves can be obtained by using the power-law expansion. The coupling among electromagnetic force and Riemann tensor ought to be appeared moderately within the region close to the super-massive dense bodies on the middle of galaxies because of their excessive mass density and very strong gravity. Generally, Weyl tensor is considered as a foremost tensor in Einstein's theory of relativity. This tensor within the spacetime explains a sort of gravitational warp. The coupling among Weyl tensor and Maxwell field may be dealt simply as a unique form of interplays between curvature tensor and electromagnetic force. Consider the fact that Weyl tensor is the combination of a curvature tensor $R_{ijuv}$, Ricci tensor $R_{ij}$, and Ricci scalar $R$.
In the view of well-known principle of Einstein theory of general relativity, photons would be deviated from their original straightforward path whenever they pass very near to the dense and heavy objects and the alternative results are said to be gravitational lensing \cite{17}-\cite{19}. The snap shots of the stars within the gravitational lensing bring the statistics around these stars and gravitational lens itself. Moreover, this statistics can assist us to discover the more dense astrophysical bodies inside the Universe and observe more ideas within the strong field. Several inspections have calculated the propagation of unfastened photon inside the spacetime and also studied the outcomes
of the spacetimes factors over the gravitational lensing \cite{17}-\cite{37}. In the electro-dynamics of the Born-Infeld, Eiroa \cite{38} also
investigated the behavior of photon and discovered that during this situation, the photon did not pursue the geodesics of the line element. However, the photon pursued the geodesics of powerful line element relying at the coupling of the Born-Infeld, whereas this coupling varies the characteristics of the gravitational lensing. Generally, the gravitational lensing must rely upon the photon and features of the past spacetime.

Consequently, it is far of curiosity to investigate how the interplay among spacetime Weyl tensor and the photon have an effect on the gravitational lensing. By the preceding conversation, we understand that the coupling among Weyl tensor and Maxwell tensor will vary the attitude of electromagnetic area inside the past spacetime. It is widely recognized that light is really a form of electromagnetic waves, this shows that the coupling shall vary the photon propagation inside the background spacetime and carry to a few precise aspects of gravitational lensing. The deflecting angle had also been investigated for the photon coupling to the curvature tensor within the weak field approximate limit in Ref.\cite{2}. Because the weak field approximate limit attains simply a $1^{st}$ array variation by the Minkowski spacetime and only it is accurate inside area away from the black holes. Also, it is important to search out further the gravitational lensing inside the strong field area close to the black hole due to the fact that it begins from whole seize of the coupling photon and prevails the special array within the diversity of the deflecting angle. Furthermore, a good way to investigate the general functions of the deflecting angle of the photon coupling to the Weyl tensor, presently we shall examine the strong gravitational lensing inside the Kiselev black hole. After this, we shall search the consequences of that coupling at the deflecting angle and other observable inside the strong field approximate limit.
The plan of the paper has been organized as follows: The next section is devoted to derive the equation of motion for the photons coupled to Weyl tensor. Sec.\textbf{3}, deals with null geodesic and equation of photon sphere. We study the strong gravitational lensing observables for Kiselev black hole spacetime in section \textbf{4}.
Finally, we summaries our results and compares these with Ref.\cite{1} in the last section.
\section{Equation of motion for the photons coupled to Weyl tensor }
This section is devoted to formulate the equation of motion for the photons coupled to Weyl tensor in $4D$ black hole solution by the geometric optics approximation \cite{2},\cite{39}-\cite{42}. The action of electromagnetic field coupled to Weyl tensor can be expressed as \cite{1,10}
\begin{eqnarray}
S = \int d^{4} x\sqrt{-g} [ \frac{R}{16\pi G} -
 \frac{1}{4} (F_{\mu\nu}F^{\mu\nu} -
4\alpha C_{\mu\nu\rho\sigma} F^{\mu\nu}F^{\rho\sigma})],
\end{eqnarray}
where $C_{\mu\nu\rho\sigma}$ represents the Weyl tensor. For $n$ dimensional spacetime it is defined by
\begin{eqnarray}
C_{\mu\nu\rho\sigma} = R_{\mu\nu\rho\sigma} -
 \frac{2}{n - 2} ( g_{\mu} [ _{\rho} R _{\sigma} ]_{\nu} - g_{\nu}
 [_{\rho} R _{\sigma}]_{\mu}) + \frac{2}{(n - 1 )(n - 2 )} R g_{\mu}
  [_{\rho} g _{\sigma} ]_{\nu}.
\end{eqnarray}
Also,  $F_{\mu\nu}$ is the
usual electromagnetic tensor given by
\begin{eqnarray}
F_{\mu\nu} = A_{\mu ; \nu} - A_{\nu ;\mu}.
\end{eqnarray}
The coupling parameter $\alpha$ has dimensions of length-squared. Varying
the action (1) with respect to $A_{\mu}$, one can obtain the following Maxwell equation with Weyl correction
\begin{eqnarray}
\nabla_{\mu} ( F_{\mu\nu} - 4\alpha C_{\mu\nu\rho\sigma}F^{\rho\sigma} ) = 0.
\end{eqnarray}
With the intention to get the generalized equation for moving photons from above mentioned equation, we apply the geometric optics
approximation \cite{2},\cite{39}-\cite{42}, according to this approximation wavelength of photon $\lambda_{p}$ is very smaller than a
regular curvature scale $\emph{L}$, but is larger than the electron Compton wavelength
$\lambda_{e}$, i.e., $ \lambda_{e} < \lambda < \emph{L} $. This guarantees that the
variation of gravitational and electromagnetic fields with the
standard curvature scale are negligible during the process of photon propagation. In the light of this approximation \cite{2},\cite{39}-\cite{42}, the electromagnetic field tensor can be defined as follows \cite{1}:
\begin{eqnarray}
F_{\mu\nu} = \emph{f}_{\mu\nu} \emph{e}^{i\theta},
\end{eqnarray}
where $\emph{f}_{\mu\nu}$ is a slowly varying amplitude and $\theta$ is a rapidly varying parameter. In this approximation
the $\emph{f}_{\mu\nu; \lambda} $ may be omitted. The wave vector is $ \emph{k}_{\mu} = \partial_{\mu} \theta $,
which may be handled in the quantum particle theory as a standard photon momenta. The amplitude $\emph{f}_{\mu\nu}$ is restricted by means of the Bianchi
identity \cite{1}
\begin{eqnarray}
D_{\lambda}F_{\mu\nu} + D_{\mu}F_{\nu\lambda} + D_{\nu}F_{\lambda\mu} = 0,
\end{eqnarray}
which leads to \cite{1}
\begin{eqnarray}
\emph{k}_{\lambda}\emph{f}_{\mu\nu} + \emph{k}_{\mu}\emph{f}_{\nu\lambda} +
\emph{k}_{\nu}\emph{f}_{\lambda \mu} = 0.
\end{eqnarray}
The amplitude $\emph{f}_{\mu\nu}$ can be written as \cite{1}
\begin{eqnarray}
\emph{f}_{\mu\nu} = \emph{k}_{\mu}\emph{a}_{\nu} - \emph{k}_{\nu}\emph{a}_{\mu},
\end{eqnarray}
where $\emph{a}_{\mu}$ is the polarization vector satisfying the condition
\begin{eqnarray}
{k}^{\mu}\emph{a}_{\mu} = 0.
\end{eqnarray}
The amplitude $\emph{f}_{\mu\nu}$
has just $3$ independent components. Using the Eqs.{(5)} and {(8)} into
Eq.{(4)}, we get the equation of motion of photon coupled to Weyl
tensor
\begin{eqnarray}
\emph{k}_{\mu}\emph{k}^{\mu}\emph{a}^{\nu} + 8\alpha C^{\mu\nu\rho\sigma}
\emph{k}_{\sigma}\emph{k}_{\mu}\emph{a}_{\rho} = 0.
\end{eqnarray}
The spherically symmetric and static Kiselev black hole spacetime is defined by \cite{z4}
\begin{eqnarray}
ds^2 = - f(r)dt^2 + f^{-1}(r) dr^2 + r^2 (d\theta^2 +  \sin^2\theta d\phi^2),
\end{eqnarray}
with
\begin{eqnarray}
f(r) = 1 - \frac{2M}{r} - \sigma r,
\end{eqnarray}
where, $\sigma$ is Kiselev parameter, $M$ and $r$ both are the
black hole mass and radius, respectively. A lot of work related to Kiselev black hole and gravitational lensing is available in literature \cite{rev1}-\cite{rev10}. According to Kiselev \cite{z4}, the spactime (11) is the static spherically symmetric solution of Einstein field equations, which represents a black hole surrounded by the quintessence field, but recently, Visser \cite{v1} have proved that Kiselev solution is neither a perfect solution nor quintessence. Although, it was mentioned in more than 200 articles that Kiselev spacetime is quintessence solution, but Visser proved that it is wrong to associate the $\textit{term}$ quintessence field with the Kiselev spacetime. After the Visser's correction about the description of Kiselev black hole, Boonserm et al. \cite{v2} have investigated that \textit{the anisotropic stress-energy leading to Kiselev black hole solution can be represented by being split into a perfect fluid component plus either an electromagnetic component or a scalar field component, thereby quantifying the precise extent to which the Kiselev black hole fails to represent a perfect fluid spacetime}.

The event horizons of a Kiselev black hole spacetime, can be obtained by
taking $g_{00} = 0$. Thus, we have
\begin{eqnarray}
r_{h+} = \frac{1 + \sqrt{1 - 8M\sigma}}{2\sigma},
\end{eqnarray}
\begin{eqnarray}
 r_{h-} =\frac{1 - \sqrt{1 - 8M\sigma}}{2\sigma},
\end{eqnarray}

where, the region $r_{h+}$ represents the outer horizon, whereas
$r_{h-}$ represents the inner horizon of the black hole, respectively.\\
For $\sigma =0$ into Eq.(12), we get just single black hole event horizon, known as Schwarzschild black
hole event horizon, i;e. $r_{h}$ = $2M$. Now, we introduce the black hole spacetime as a set of orthonormal frames , for this, we have the field of
vierbeins defined by \cite{1}
\begin{eqnarray}
g_{\mu\nu} = \eta_{ab}\emph{e}^{a}_{\mu}\emph{e}^{b}_{\nu},
\end{eqnarray}
For instance equation (11) for the vierbeins should read as follows:
\begin{eqnarray}
\emph{e}^{a}_{\mu} = diag( \sqrt{f}, \frac{1}{\sqrt{f}}, r, rsin\theta ),
\end{eqnarray}
and its inverse is
\begin{eqnarray}
\emph{e}^{\mu}_{a} = diag( \frac{1}{\sqrt{f}}, \sqrt{f}, \frac{1}{r}, \frac{1}{rsin\theta} ),
\end{eqnarray}
where $\eta_{ab}$ defines the Minkowski metric and $\emph{e}^{a}_{\mu}, \emph{e}^{b}_{\nu}$ are called  vierbeins (or tetrads).
In terms of antisymmetric bi-tensor, vierbeins can be written as \cite{1,2,39}:
\begin{eqnarray}
U^{ab}_{\mu\nu} = \emph{e}^{a}_{\mu}\emph{e}^{b}_{\nu} - \emph{e}^{a}_{\nu}\emph{e}^{b}_{\mu}.
\end{eqnarray}
Here, the Weyl tensor may be rewritten as follows \cite{1}:
\begin{eqnarray}
C_{\mu\nu\rho\sigma} = \mathcal{A}( 2U^{01}_{\mu\nu}U^{01}_{\rho\sigma} -
 U^{02}_{\mu\nu}U^{02}_{\rho\sigma} -U^{03}_{\mu\nu}U^{03}_{\rho\sigma} +
  U^{12}_{\mu\nu}U^{12}_{\rho\sigma} + U^{13}_{\mu\nu}U^{13}_{\rho\sigma} -
    2U^{23}_{\mu\nu}U^{23}_{\rho\sigma} ),
\end{eqnarray}
where
\begin{eqnarray}
\mathcal{A} = - \frac{1}{12r^{2}} [r^2 f'' - 2f'r + 2f - 2  ].
\end{eqnarray}
In order to drive the equation of motion \cite{1,2,39}, we can introduce three combinations of momentum components \cite{1}:
\begin{eqnarray}\nonumber
l_{\nu} = \emph{k}^{\mu}U^{01}_{\mu\nu},
\end{eqnarray}
\begin{eqnarray}\nonumber
n_{\nu} = \emph{k}^{\mu}U^{02}_{\mu\nu},
\end{eqnarray}
\begin{eqnarray}
m_{\nu} = \emph{k}^{\mu}U^{23}_{\mu\nu}.
\end{eqnarray}
With the dependent combinations \cite{1}:
\begin{eqnarray}\nonumber
p_{\nu} = \emph{k}^{\mu}U^{12}_{\mu\nu} = \frac{1}{\emph{k}^{0}}(\emph{k}^{1}n_{\nu} - \emph{k}^{2}l_{\nu}),
\end{eqnarray}
\begin{eqnarray}\nonumber
r_{\nu} = \emph{k}^{\mu}U^{03}_{\mu\nu} =  \frac{1}{\emph{k}^{2}}(\emph{k}^{0}m_{\nu} + \emph{k}^{3}l_{\nu}),
\end{eqnarray}
\begin{eqnarray}
q_{\nu} = \emph{k}^{\mu}U^{13}_{\mu\nu} =   \frac{\emph{k}^{1}}{\emph{k}^{0}}m_{\nu} +
\frac{\emph{k}^{1}\emph{k}^{3}}{\emph{k}^{2}\emph{k}^{0}}n_{\nu} - \frac{\emph{k}^{3}}{\emph{k}^{0}}l_{\nu}.
\end{eqnarray}
The polarization vectors $ l_{\nu}$, $ n_{\nu}$, $m_{\nu}$ are independent and orthogonal to the wave vector $\emph{k}_{\nu}$. Making use of the relation (22)
and contract the equation $(9)$ with respect to $ l_{\nu}$, $ n_{\nu}$ and
$m_{\nu}$, respectively. One can obtain
\begin{eqnarray}\nonumber
0 = \emph{k}^{2}\emph{a}.\emph{l} &+& 16\alpha\emph{W}( \emph{l}^{2}\emph{a}.\emph{l} - \emph{l}.m\emph{a}.m)  \\&-&
8\alpha\emph{W}(\emph{l}.n\emph{a}.n + \emph{l}.r\emph{a}.r -
  \emph{l}.p\emph{a}.p - \emph{l}.q\emph{a}.q ).
\end{eqnarray}
Similarly, other two equations are
\begin{eqnarray}\nonumber
0 = \emph{k}^{2}\emph{a}.n &+& 16\alpha\emph{W}( n.\emph{l}\emph{a}.\emph{l} -  n.m\emph{a}.m) \\ &-&
8\alpha\emph{W}( n^{2}\emph{a}.n +  n.r\emph{a}.r -
 n.p\emph{a}.p -  n.q\emph{a}.q ),
\end{eqnarray}
and
\begin{eqnarray}\nonumber
0 = \emph{k}^{2}\emph{a}.m &+& 16\alpha\emph{W}(m.\emph{l}\emph{a}.\emph{l} - m^{2}\emph{a}.m) \\&-&
8\alpha\emph{W}(m.n\emph{a}.n +  m.r\emph{a}.r -
  m.p\emph{a}.p -  m.q\emph{a}.q ).
\end{eqnarray}
Matrix form,

\begin{eqnarray}
\left(
  \begin{array}{ccc}
    K_{11} & 0 & 0 \\
    K_{21} & K_{22} & K_{23} \\
    0 &    0 & K_{33} \\
  \end{array}
\right)
\left(
  \begin{array}{ccc}
   a  .l  \\
   a  .n  \\
   a  .m  \\
  \end{array}
\right)
= 0.
\end{eqnarray}
\begin{eqnarray}
K_{11} =  (1 + 16 \alpha \mathcal {A})(-(e^{0}_{t})^{2}\emph{k}^{0}\emph{k}^{0} + (e^{1}_{r})^{2}\emph{k}^{1}\emph{k}^{1}) + (1 - 8 \alpha \mathcal {A})( (e^{2}_{\theta})^{2} \emph{k}^{2}\emph{k}^{2} + (e^{3}_{\phi})^{2} \emph{k}^{3}\emph{k}^{3}),
\end{eqnarray}
\begin{eqnarray}
K_{21} =  24 \alpha \mathcal {A} \sqrt{(e^{1}_{r})^{2}(e^{2}_{\theta})^{2}}\emph{k}^{1}\emph{k}^{2} , && K_{23} =  - 24 \alpha \mathcal {A} \sqrt{-(e^{0}_{t})^{2}(e^{3}_{\phi})^{2}} \emph{k}^{0}\emph{k}^{3} ,
\end{eqnarray}
\begin{eqnarray}
K_{22} =  (1 - 8 \alpha \mathcal {A})(-(e^{0}_{t})^{2}\emph{k}^{0}\emph{k}^{0} + (e^{1}_{r})^{2}\emph{k}^{1}\emph{k}^{1} + (e^{2}_{\theta})^{2} \emph{k}^{2}\emph{k}^{2} + (e^{3}_{\phi})^{2} \emph{k}^{3}\emph{k}^{3}),
\end{eqnarray}
\begin{eqnarray}
K_{33} =  (1 - 8 \alpha \mathcal {A}) (-(e^{0}_{t})^{2}\emph{k}^{0}\emph{k}^{0} + (e^{1}_{r})^{2}\emph{k}^{1}\emph{k}^{1}) + (1 + 16 \alpha \mathcal {A})( (e^{2}_{\theta})^{2} \emph{k}^{2}\emph{k}^{2} +  (e^{3}_{\phi})^{2} \emph{k}^{3}\emph{k}^{3}).
\end{eqnarray}

These coefficients can be reduced to the following form:

\begin{eqnarray}
K_{11} = ( 1 + 16\alpha \mathcal {A} )(
g_{00}\emph{k}^{0}\emph{k}^{0} + g_{11}\emph{k}^{1}\emph{k}^{1} )  +  (1 -
8\alpha \mathcal {A}) ( g_{22}\emph{k}^{2}\emph{k}^{2} +
g_{33}\emph{k}^{3}\emph{k}^{3}),
\end{eqnarray}
\begin{eqnarray}
K_{22} = ( 1 - 8\alpha \mathcal {A})( g_{00}\emph{k}^{0}\emph{k}^{0} +
g_{11}\emph{k}^{1}\emph{k}^{1}   +  g_{22}\emph{k}^{2}\emph{k}^{2} +
g_{33}\emph{k}^{3}\emph{k}^{3}),
\end{eqnarray}
\begin{eqnarray}
K_{21} = 24\alpha \mathcal {A}\sqrt{g_{11}g_{22}}\emph{k}^{1}\emph{k}^{2},
K_{23} = -24\alpha \mathcal {A}\sqrt{g_{00}g_{33}}\emph{k}^{0}\emph{k}^{3},
\end{eqnarray}
\begin{eqnarray}
K_{33} = ( 1 - 8\alpha \mathcal {A})( g_{00}\emph{k}^{0}\emph{k}^{0} +
g_{11}\emph{k}^{1}\emph{k}^{1} )  +  (1 + 16\alpha \mathcal {A})
( g_{22}\emph{k}^{2}\emph{k}^{2} + g_{33}\emph{k}^{3}\emph{k}^{3}).
\end{eqnarray}

The possibility of Eq.(26) is given as $K_{11}K_{22}K_{33} =
0$ with a nonzero solution. The first root $K_{11} = 0$ results to the modified light cone
\begin{eqnarray}
(1 + 16\alpha \mathcal {A} )(g_{00}\emph{k}^{0}\emph{k}^{0} +
 g_{11}\emph{k}^{1}\emph{k}^{1}) + (1 - 8\alpha \mathcal {A})(g_{22}\emph{k}^{2}
 \emph{k}^{2} + g_{33}\emph{k}^{3}\emph{k}^{3}) = 0,
\end{eqnarray}
in which both  the polarization vector $\emph{a}_{\mu}$ and momentum component $\emph{l}_{\mu}$ are proportional to each other and the strength
\begin{eqnarray}
\emph{f}_{\mu\nu} \varpropto (\emph{k}_{\mu}l_{\nu} - \emph{k}_{\nu}l_{\mu}).
\end{eqnarray}
The second root $K_{22} = 0$ signifies that $\emph{a} .l = 0 = \emph{a} .m$ in Eq.(26). This root suggests that $\emph{a}_{\mu} = \lambda \emph{k}_{\mu}$
and $\emph{f}_{\mu\nu}$ vanishes \cite{2}. However, the second root will correspond
to the unphysical polarisation, while, the second root must be unnoticed for the standard propagating directions of the coupling photon. The third root, $K_{33} =
0$, i.e.,
\begin{eqnarray}
(1 - 8\alpha \mathcal {A})(g_{00}\emph{k}^{0}\emph{k}^{0} +
 g_{11}\emph{k}^{1}\emph{k}^{1}) + (1 + 16\alpha \mathcal {A})(g_{22}\emph{k}^{2}
 \emph{k}^{2} + g_{33}\emph{k}^{3}\emph{k}^{3}) = 0,
\end{eqnarray}
which means that the vector
\begin{eqnarray}
\emph{a}_{\mu} = \lambda\emph{m}_{\mu},
\end{eqnarray}
and the strength
\begin{eqnarray}
\emph{f}_{\mu\nu} = \lambda (\emph{k}_{\mu}m_{\nu} - \emph{k}_{\nu}m_{\mu}).
\end{eqnarray}
The light cone conditions depend on the photons coupled to Weyl tensor as well as polarization directions. Further, for the coupling
photon, the consequences of Weyl tensor eventually different with the several polarization on the propagation of photons and in the spacetime  of \cite{39}-\cite{42} these consequences provide a development of
birefringence. As the parameter of coupling $\alpha$ is equal to zero, the light-cone conditions (35) and (37) obtain again to the standard shape
without Weyl corrections. Now, we assist the Eq.(20) for a Kiselev black hole spacetime ,
the light cone conditions (35) and (37) can be rewritten as follows:
\begin{eqnarray}
(1 + \frac{16\alpha M}{r^{3}} )(g_{00}\emph{k}^{0}\emph{k}^{0} +
 g_{11}\emph{k}^{1}\emph{k}^{1}) + (1 - \frac{8\alpha M}{r^{3}})(g_{22}\emph{k}^{2}
 \emph{k}^{2} + g_{33}\emph{k}^{3}\emph{k}^{3}) = 0,
\end{eqnarray}
\begin{eqnarray}
(1 -  \frac{8\alpha M}{r^{3}} )(g_{00}\emph{k}^{0}\emph{k}^{0} +
 g_{11}\emph{k}^{1}\emph{k}^{1}) + (1 + \frac{16\alpha M}{r^{3}})(g_{22}\emph{k}^{2}
 \emph{k}^{2} + g_{33}\emph{k}^{3}\emph{k}^{3}) = 0.
\end{eqnarray}
The relation (40) represents the light cone condition along the polarization
vector $ l_{\mu}$ for the coupled photon (PPL), whereas the relation(41) is the light
cone condition along the polarization vector $ m_{\mu} $ for the coupled photon (PPM), respectively.

\section{Null Geodesic and Equation of Photon Sphere}
The light cone conditions (40)
and (41) show that the photons coupled to Weyl tensor follow null geodesics of the effective metric  $ \gamma_{\mu\nu} $ \cite{1}, i.e;
\begin{eqnarray}
\gamma^{\mu\nu}\emph{k}_{\mu}\emph{k}_{\nu} = 0.
\end{eqnarray}
The effective metric
can be defined as follows:
\begin{eqnarray}
ds^{2} = - A(r)dt^{2} + B(r)dr^{2} + C(r)
W(r)^{-1}(d\theta^{2} + sin^{2}\theta d\phi^{2}),
\end{eqnarray}
where the metric
functions $A(r), ~B(r), ~C(r)$ and $W(r)$ are given by
\begin{eqnarray}
A(r) = 1- \frac{2M}{r} - \sigma r
\end{eqnarray}
\begin{eqnarray}
B(r) = \frac{1}{A(r)}
\end{eqnarray}
\begin{eqnarray}
C(r) = r^{2},
\end{eqnarray}
\begin{eqnarray}
W(r) = \frac{r^{3} - 8\alpha M}{r^{3} + 16\alpha M },
\end{eqnarray}
for PPL case and
\begin{eqnarray}
W(r) = \frac{r^{3} + 16\alpha M}{r^{3} - 8\alpha M},
\end{eqnarray}
for PPM case, respectively. The metric functions are dependable
functions of the photon polarization directions. Under the possibility $(\theta =
\frac{\pi}{2})$
 the effective metric (43) can be reduced to the following form
\begin{eqnarray}
ds^{2} = - A(r)dt^{2} + B(r)dr^{2} + C(r)W(r)^{-1}d\phi^{2}.
\end{eqnarray}
For the equatorial plane $(\theta =
\frac{\pi}{2}, \emph{k}_{\theta} = 0)$, the wave vector becomes $
\emph{k}_{\mu} = (\emph{k}_{0}, \emph{k}_{1}, 0, \emph{k}_{3})$ with $ \emph{k}_{2} = 0$ and the
simplification of the polarisation vectors $m_{\mu}$ and $l_{\mu}$ can be
more formed as
\begin{eqnarray}
m_{\mu} = (0, 0, -\emph{k}^{3}, 0), l_{\mu} = (-\emph{k}^{1}, \emph{k}^{0}, 0, 0).
\end{eqnarray}
Eq.(50) shows that the polarization vector $l_{\mu}$ is
situated on the equatorial plane, whereas the polarization vector $m_{\mu}$
indicates the polarization which is perpendicular to the equatorial plane of
motion, respectively.
When $\sigma\rightarrow0$, we obtain the effective metric given in \cite{1}.
The coupled photons trajectory has become limited on the
equatorial plane. So, using the
condition $(g_{\mu\nu}u^{\mu}u^{\nu} = 0)$ for the four velocity $u^{\mu}$ . We attain the null geodesic's equation for the coupling photon in a Kiselev  space-time.
\begin{eqnarray}
(\frac{dr}{d\lambda})^2 = \frac{1}{B(r)}\left(\frac{E^{2}}{A(r)}  - W(r)\frac{L^{2}}{C(r)}\right),
\end{eqnarray}
where $\lambda$ acts like a affine parameter along the null geodesic. The parameters E and L represent the energy and the angular momentum
per unit mass respectively. They are expressed as follows:
\begin{eqnarray}
E = A(r)\dot{t},&& L = C(r)W(r)^{-1}\dot{\phi}.
\end{eqnarray}
By working with the photon sphere equation \cite{20}. One can obtain the impact
parameter $u(r)$ and the equation of photon sphere
\begin{eqnarray}
u(r) = \sqrt{\frac{C(r)}{A(r)W(r)}},
\end{eqnarray}
\begin{eqnarray}
W(r)[A'(r)C(r) - A(r)C'(r)] + A(r)C(r)W'(r) = 0.
\end{eqnarray}
In a $4D$ space-time, the biggest real roots of equation (54) can be characterized as the photon sphere radius $r_{ps}$ outside the event horizon. However, it is not easy to get the analytical form of the photon sphere radius $r_{ps}$, due to the complexity of coupled term associated to Weyl tensor in Eq.(54).
To avoid such complexity problems and to obtain the radius $r_{ps}$ for coupled photons, we apply the numerical methods. Our outcomes demonstrate that the radius $r_{ps}$ of photon sphere only occurs in the system when  $\alpha_{c1}$ $>\alpha$ $>\alpha_{c2}$  both for PPM and PPL cases, respectively. Whereas for the coupled photons, the critical values can be resolved by the overlapping situation of the radius $r_{ps}$ with the event horizon. Moreover, there are critical values $\alpha_{c1}$ for PPM and $\alpha_{c2}$ for PPL which
depend on the Kiselev parameter $\sigma$ and are defined as follows:
\begin{eqnarray}
\alpha_{c1} =-2\alpha_{c2} =
\frac{1}{8}[\frac{1+3(1-8\sigma M)^{\frac{1}{2}}+3(1-8\sigma M)+(1-8\sigma M)^{\frac{3}{2}}}{8\sigma^{3}M}].
\end{eqnarray}
By setting $E = 1$, as the coupling parameter $\alpha\rightarrow0$, we find that the function $W \rightarrow 1$, which results
in that the impact parameter and the equation of circular photon orbits for PPL are the same as those for
PPM. This means that gravitational lensing is independent of the polarization directions of the photon in the
case without the coupling.
From Fig.1, it is clear that the critical value $\alpha_{c1}$ decreases and the critical value $\alpha_{c2}$ increases when $\sigma$ increases for both PPM and PPL, respectively. Applying the numerical method, we present the dependence of the photon sphere
radius $r_{ps}$ on the coupling parameter  $\alpha$ and the Kiselev parameter $\sigma$ for PPM and
PPL, as shown in Figs.2 and 3. We find that with the increase of $\alpha$, the radius $r_{ps}$ increases for PPL while
decreases for PPM. Meanwhile, when $\sigma$ increases the radius $r_{ps}$ decreases for PPL and
increases for PPM.
\\From Figs.2 and 3, we see that the features of gravitational lensing are completely different for PPL and PPM cases, respectively. In different scene, gravitational lensing depends on the photon polarization drections with the coupling parameter and the Kiselev parameter.
\begin{figure}
\includegraphics[width=80mm]{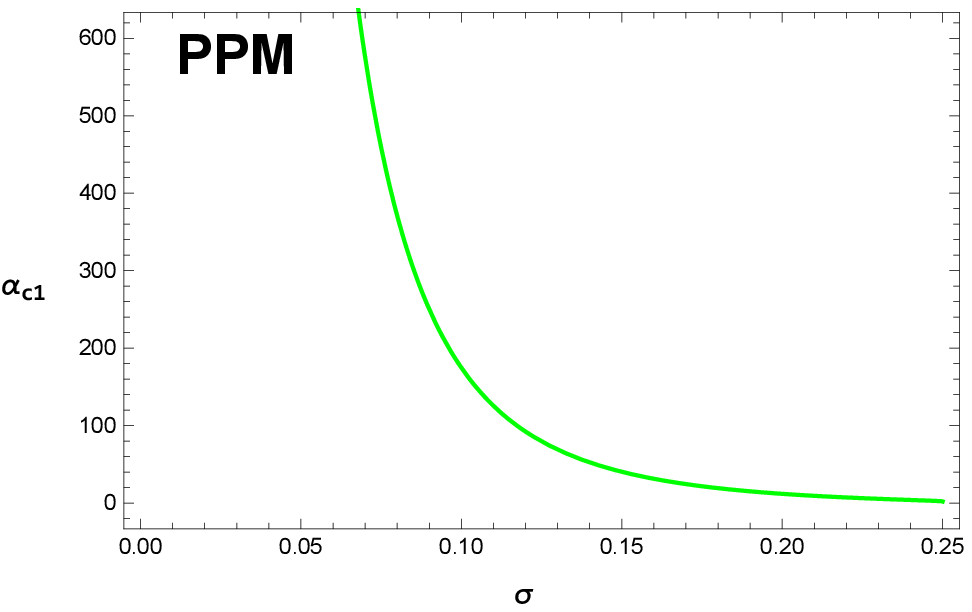}
\includegraphics[width=80mm]{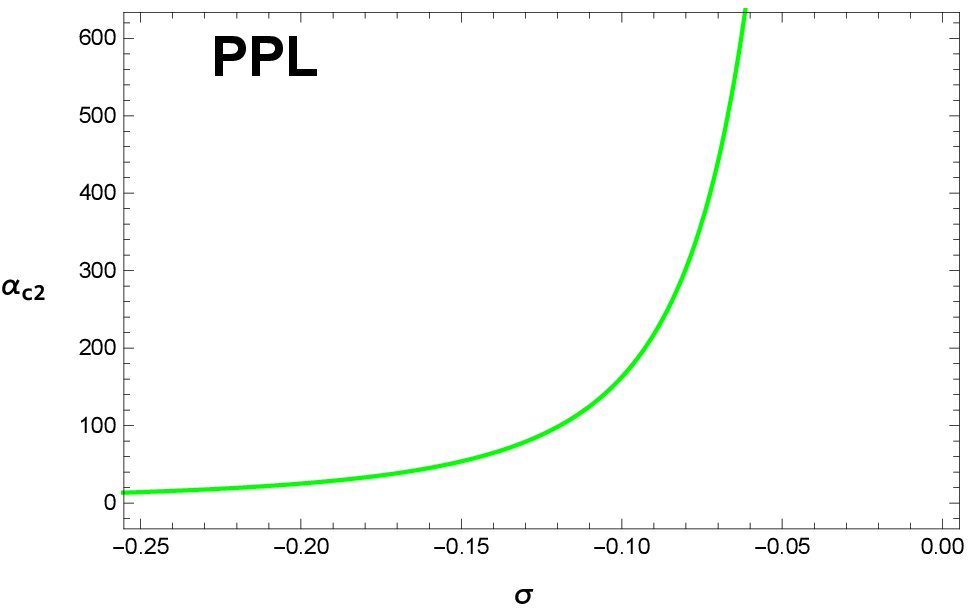}
\caption{Variation of critical values with Kiselev parameter $\sigma$ at $M = 0.5$. The photon sphere radius occurs only in the system when $\alpha$ $<\alpha_{c1}$ for PPM and $\alpha$ $>\alpha_{c2}$ for PPL, respectively.}
\end{figure}
\begin{figure}
\includegraphics[width=92mm]{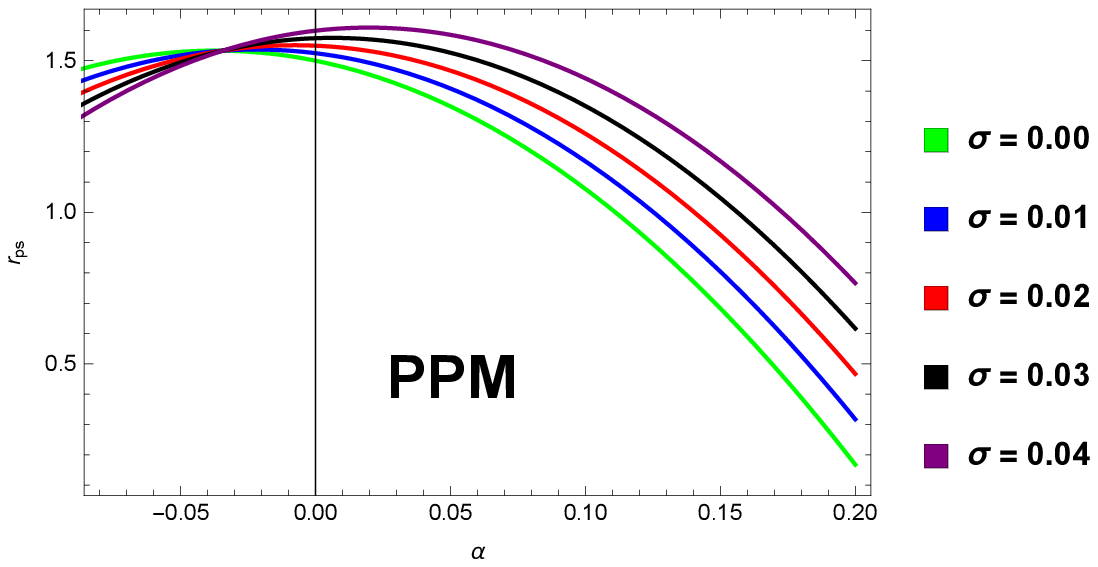}
\includegraphics[width=92mm]{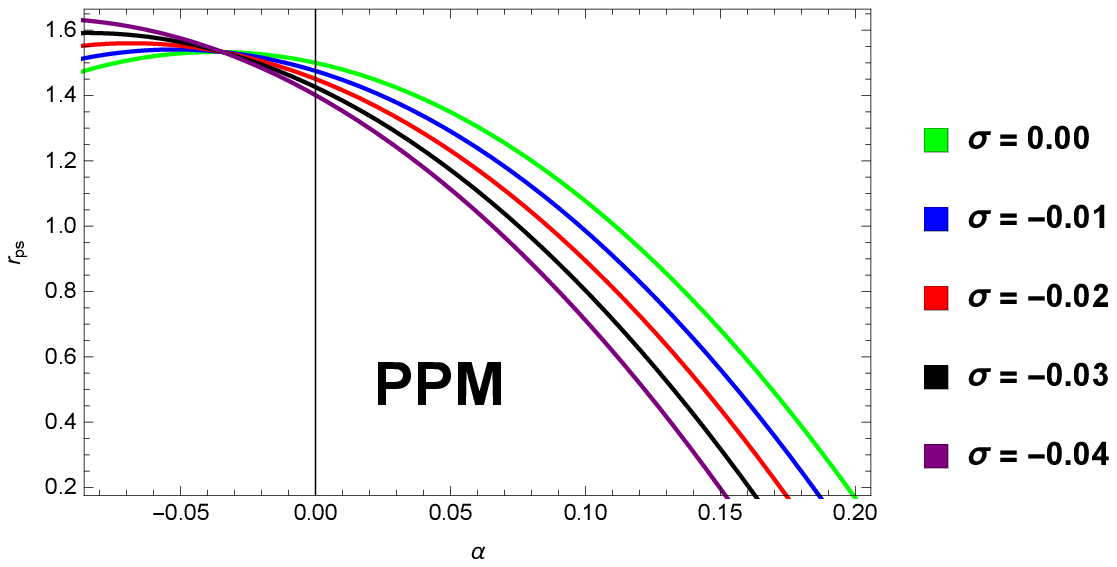}
\caption{Variation of photon sphere radius for PPM with coupling parameter $\alpha$ for different Kiselev parameter $\sigma$, where $M = 0.5$.}
\end{figure}
\begin{figure}
\includegraphics[width=92mm]{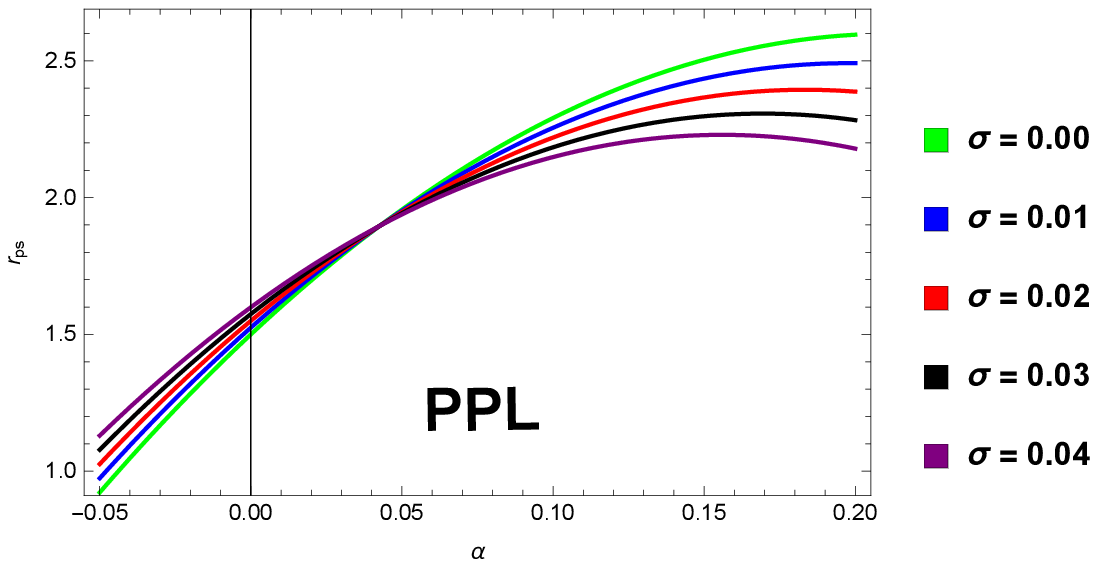}
\includegraphics[width=92mm]{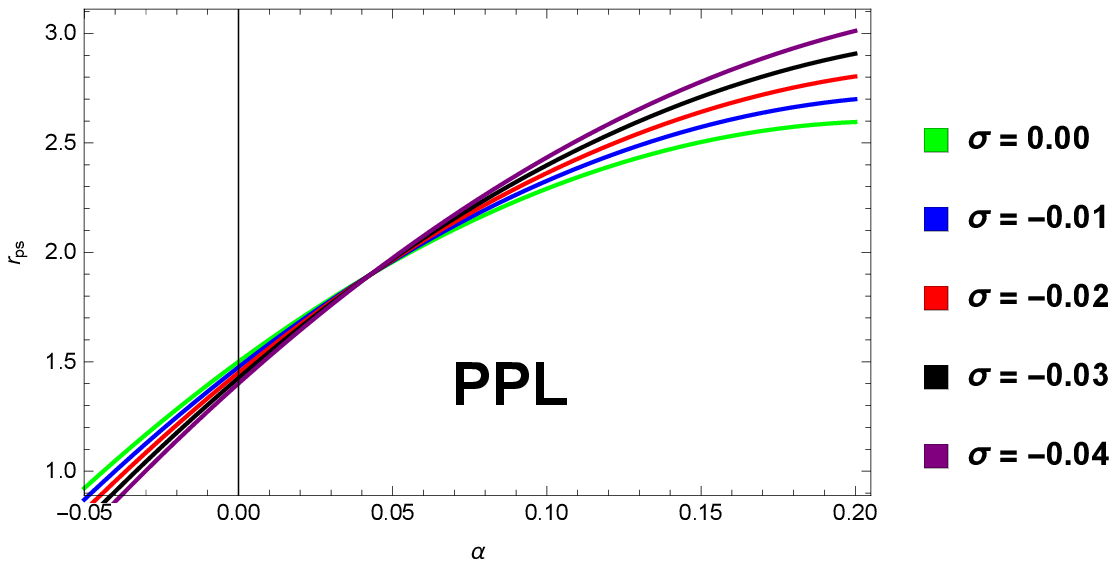}
\caption{Variation of photon sphere radius for PPL with coupling parameter $\alpha$ for different Kiselev parameter $\sigma$, where $M = 0.5$.}
\end{figure}
\section{Strong Gravitational Lensing Observables in a Kiselev Black Hole Spacetime}
Now, we discuss the following gravitational lensing observables in a Kieslev black Hole spacetime.
\subsection{Angle of Deflection}
For a photon coupled to Weyl tensor coming from infinite the relation (51) limited on equatorial
plane, one can find that the deflection angle in a Kiselev black hole spacetime is similar to that in
the case without coupling \cite{21}
\begin{eqnarray}
\alpha(r_{0}) = I(r_{0}) - \pi,
\end{eqnarray}
with
\begin{eqnarray}
I(r_{0}) = 2\int_{r_{0}}^{\infty} \frac{dr}{\sqrt{\frac{A(r)C(r)}{W(r)}}\sqrt{\frac{C(r)A(r_{0})W(r_{0})}{A(r)W(r)C(r_{0})}-1}},
\end{eqnarray}
where the variable $r_{0}$ is the closest approach distance, while $I(r_{0})$ depends on the polarization directions of photons coupled to Weyl tensor in a background spacetime. This implies that the physical properties of the deflection angle for PPM is different from that of PPL. Moreover, we can use the method of approximation proposed by Bozza \cite{27}, which helps us to study the analytic features of the angle of deflection.  For this, we have a new variable \cite{z8}
\begin{eqnarray}
z = 1 - \frac{r_{0}}{r},
\end{eqnarray}
So, the Eq.(57) yields
\begin{eqnarray}
I(r_{0}) = \int_{0}^{1} R(z,r_{0})F(z,r_{0})dz,
\end{eqnarray}
where
\begin{eqnarray}
R(z,r_{0}) = 2\frac{W(r)r^{2}\sqrt{C(r_{0})}}{r_{0}C(r)} = 2W(z,r_{0}),
\end{eqnarray}
\begin{eqnarray}
F(z,r_{0}) = \frac{1}{\sqrt{A(r_{0})W(r_{0})-\frac{A(z,r_{0})W(z,r_{0})C(r_{0})}{C(z,r_{0})}}}.
\end{eqnarray}

For all values of $r_{0}$ and z, the functions $R(z,r_{0})$ are regular. Similarly, when $z\rightarrow 0$, the other functions $F(z,r_{0})$ are divergent. Due to this reason, the integral (59) may be separated into two different sorts. One is divergent $I_{D}(r_{0})$ and the other part is regular $I_{R}(r_{0})$ with different polarizations,i.e.,
\begin{eqnarray}
I_{D}(r_{0}) = \int_{0}^{1} R(0,r_{ps})F_{0}(z,r_{0})dz,
\end{eqnarray}
\begin{eqnarray}
I_{R}(r_{0}) = \int_{0}^{1} [R(z,r_{o})F(z,r_{0}) - R(0,r_{ps})F_{0}(z,r_{0})]dz,
\end{eqnarray}
whereas, the new function $F_{0}(z,r_{0})$  in (63), can be obtained by expanding the
argument of the square root in  $F_{0}(z,r_{0})$  to the second order in $z$ as follows:
\begin{eqnarray}
F_{0}(z,r_{0}) = \frac{1}{\sqrt{p(r_{0})z + q(r_{0})z^{2}}},
\end{eqnarray}
with
\begin{eqnarray}
p(r_{0}) = - \frac{r_{0}}{C(r_{0})} \{W(r_{0})[A'(r_{0})C(r_{0}) - A(r_{0})C'(r_{0})] + A(r_{0})C(r_{0})W'(r_{0})\},
\end{eqnarray}
\begin{eqnarray}\nonumber
q(r_{0}) &=& \frac{r_{0}}{2C^{2}(r_{0})} \{2[C(r_{0}) - r_{0} C'(r_{0})][A(r_{0})W(r_{0})C'(r_{0}) - C(r_{0})(A(r_{0})W(r_{0}))']
\\&+& r_{0}C(r_{0})[A(r_{0})W(r_{0})C''(r_{0}) - C(r_{0})(A(r_{0})W(r_{0}))'']\}.
\end{eqnarray}
When the coefficient $p(r_{0})$ is nonzero $(r_{0} \neq r_{ps})$, the divergence order in $F_{0}(z,r_{0})$ is $1/\sqrt{z}$ and it may be integrated to get the possible result. When $p(r_{0})$ is zero $(r_{0} = r_{ps})$, the divergence becomes $1/z$, that originates the integral as diverge. Hence, this shows that each photon which is captured by the central object, must have $r_{0} < r_{ps}$. So, in this way, the photon can not be emerged back \cite{27}. This implies that in the strong gravitational limit field, as the photon is near to the photon sphere, the deflection angle diverges logarithmically for the coupled photons \cite{27}. Hence
\begin{eqnarray}
\alpha(\theta) = - \bar{a} \log [\frac{\theta D_{1}}{u(r_{ps})} - 1] + \bar{b}
+ O[u - u(r_{ps})],
\end{eqnarray}
with
\begin{eqnarray}\nonumber
\bar{a} = \frac{R(0,r_{ps})}{2\sqrt{q(r_{ps})}}, \\\nonumber b_{R} = I_{R}(r_{ps}),
\\\bar{b} = - \pi + b_{R} + \bar{a} \log[\frac{2r_{hs}^2 u(r_{ps})''}{u(r_{ps})}],
\end{eqnarray}

where $D_{1}$ indicates the distance between the gravitational lens and the observer. The angle $\theta = u/D_{1}$ is defined as the angular separation between the image and the lens \cite{1}. Using the relations
(47) and (48) into (68), one can find the coefficients ($\bar{a}$ and $\bar{b}$) in the strong gravitational lensing
formula (67). The variation of the functions ($\bar{a}$ and $\bar{b}$) for the coupled photon with the coupling
parameter $\alpha$ for different Kiselev parameter $\sigma$  are shown in Figs.4-5. Moreover, from
relations (67)-(68), we can study the physical properties of strong gravitational lensing for the
coupled photon in a Kiselev black hole spacetime. It is shown that both coefficients ($\bar{a}$ and $\bar{b}$)
depend not only on the polarization directions of the photon coupling with Weyl tensor, but also on the Kiselev parameter.  In Figs.4-5, we plot the variation of the functions $(\bar{a}$ and $\bar{b})$ as a numerically approach with the coupling parameter for different Kiselev parameter.
The function $\bar{a}$ constantly increases with the increase of coupling parameter and Kiselev parameter for the case of PPM, while the function $\bar{a}$ decreases for PPL with the increase of coupling parameter and increases when Kiselev parameter increases, as shown in Fig.4.
\begin{figure}
\includegraphics[width=90mm]{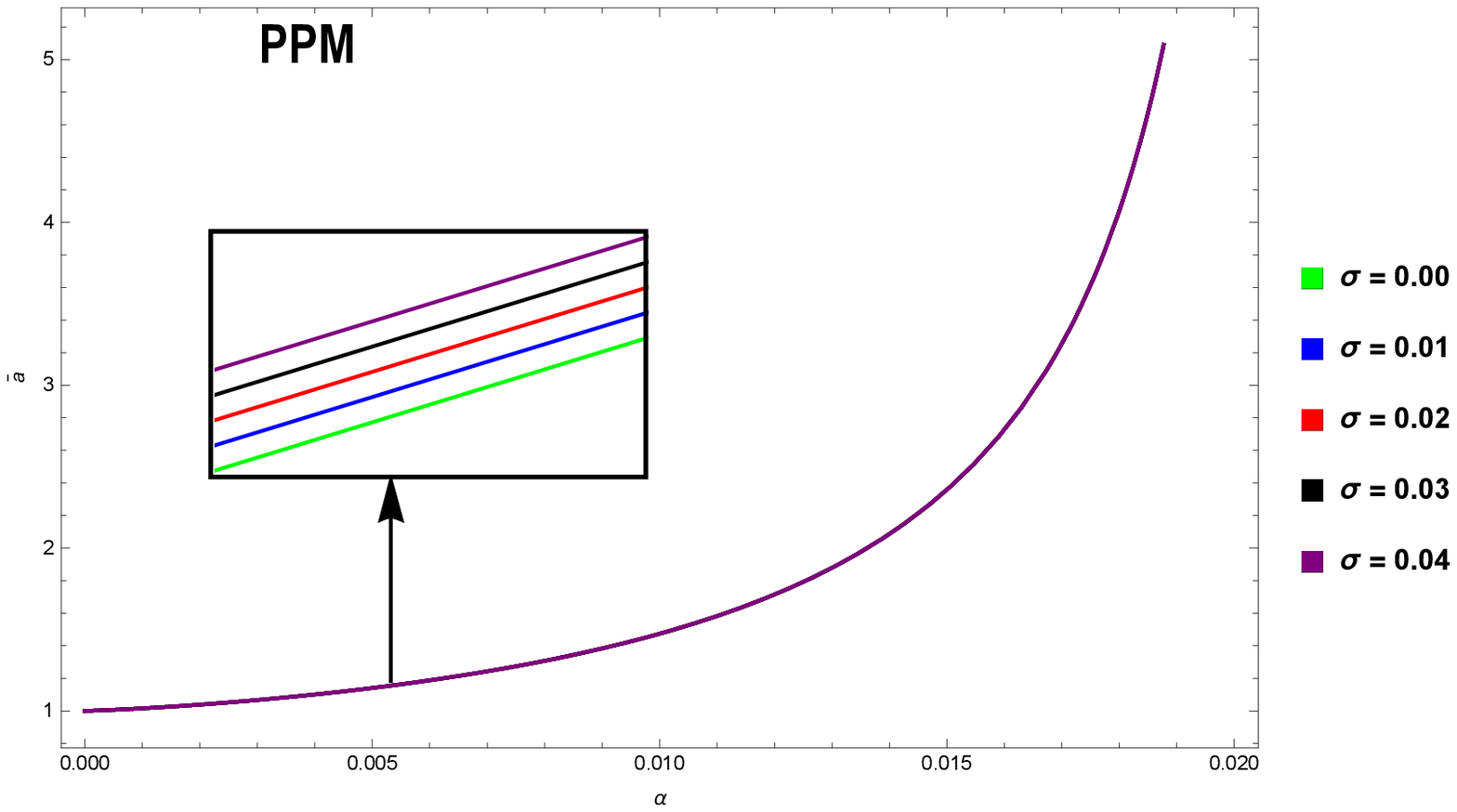}
\includegraphics[width=90mm]{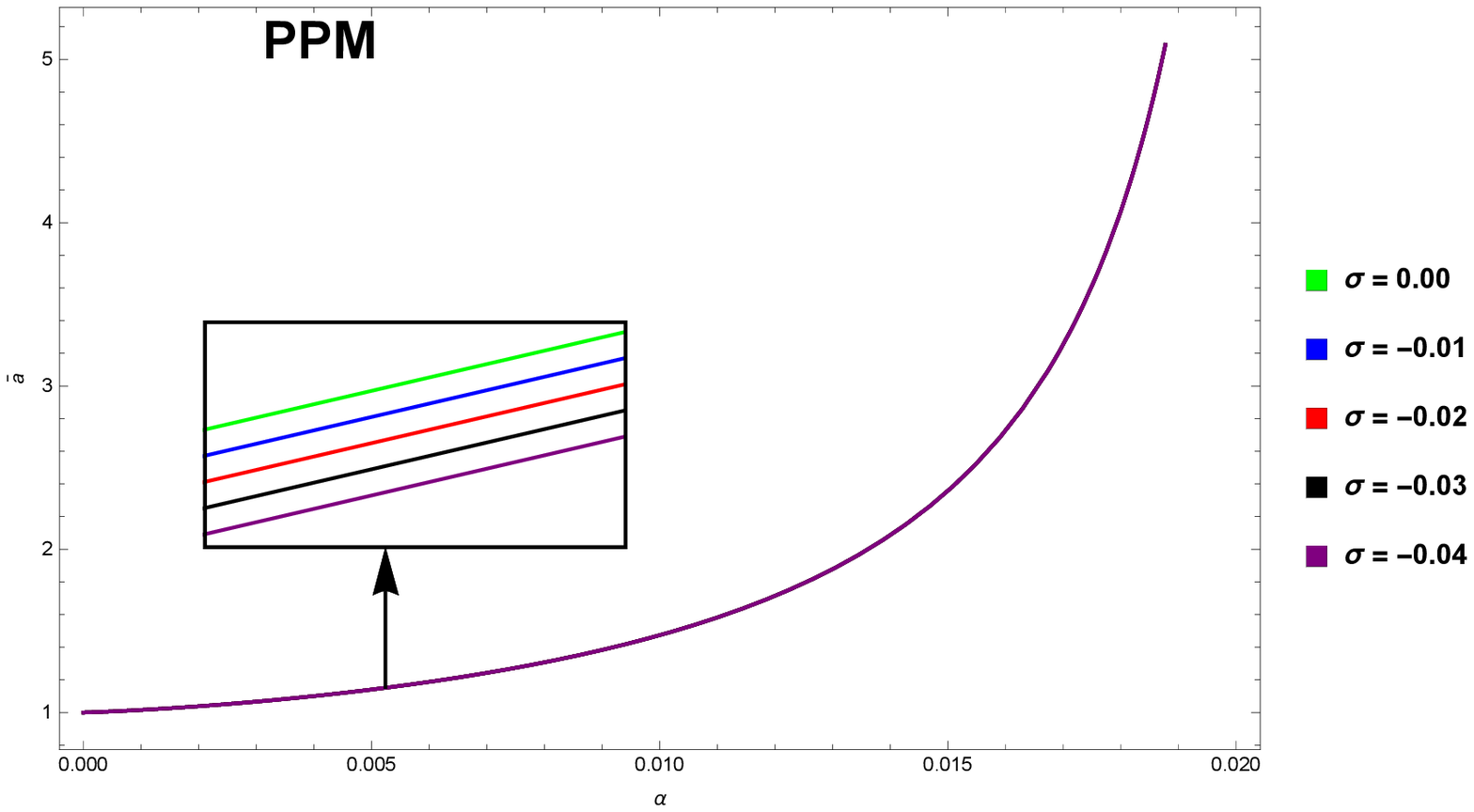}
\includegraphics[width=90mm]{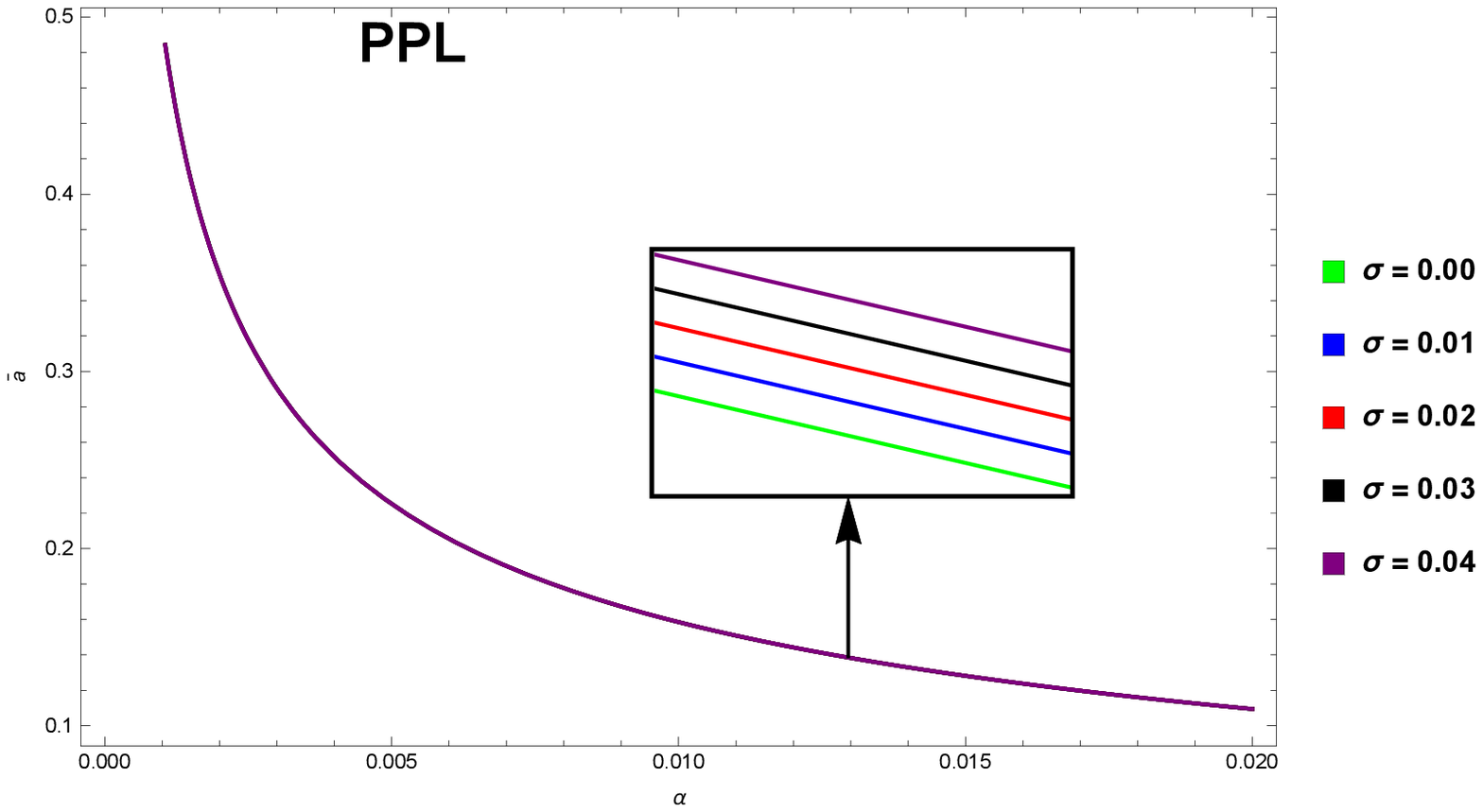}
\includegraphics[width=90mm]{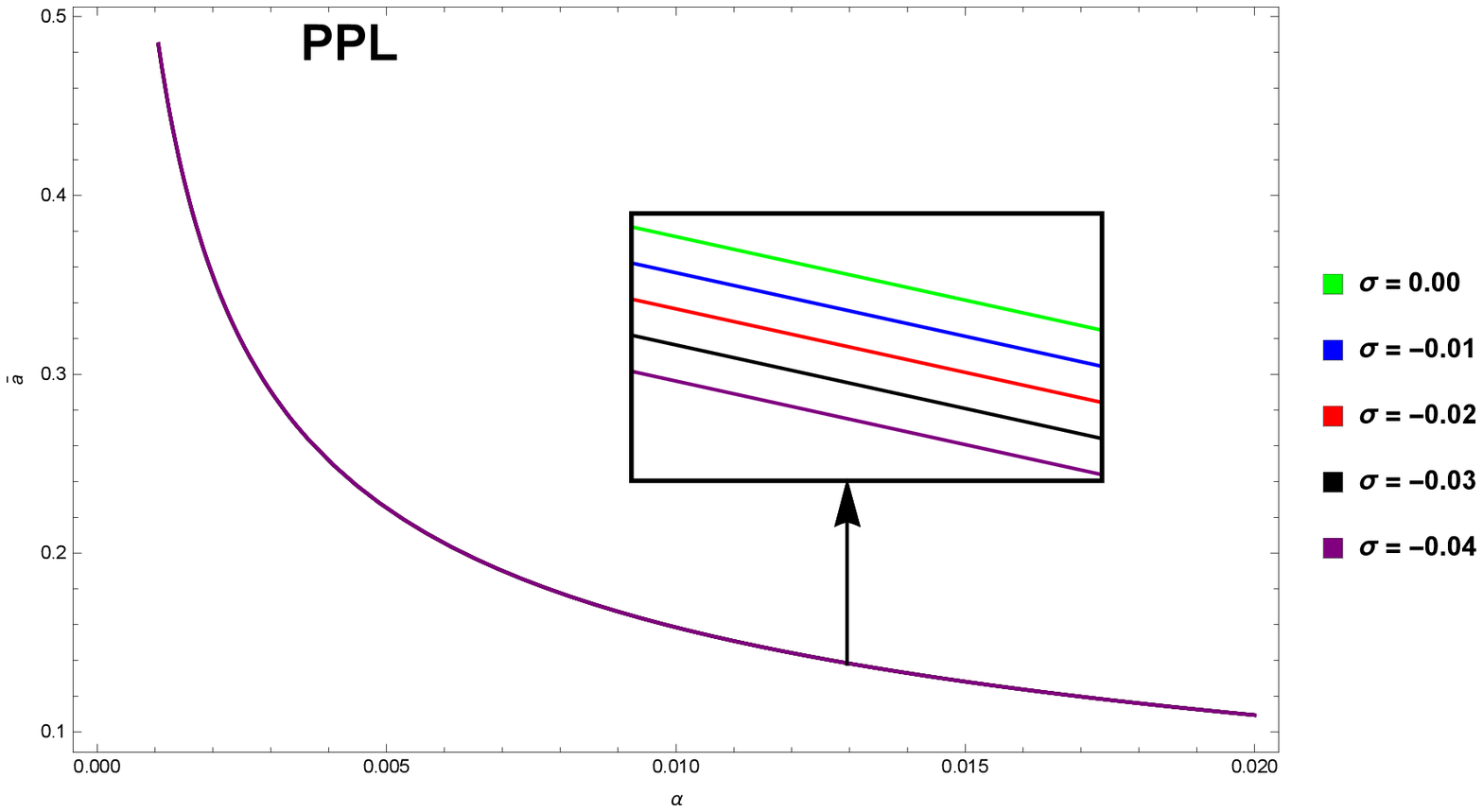}
\caption{Variation of strong deflection limit function $\bar{a}$ with coupling parameter $\alpha$ for different Kiselev parameter $\sigma$ for PPM and PPL cases, where $M = 0.5$.}
\end{figure}
The variation of $\bar{b}$ is more complex with  coupling parameter $\alpha$ for different Kiselev parameter $\sigma$. For PPM, the function $\bar{b}$ first decreases up to its minimum with the coupling parameter $\alpha$ for different values of $\sigma$ and then increases up to its maximum with the further increase of  coupling parameter $\alpha$; after that, it decreases again with  coupling parameter $\alpha$. Moreover, the variation of $\bar{b}$ with  coupling parameter $\alpha$ for the different values of Kiselev parameter $\sigma$ for the case of  PPL is totally converse to that for PPM as shown in Fig.5.
\begin{figure}
\includegraphics[width=90mm]{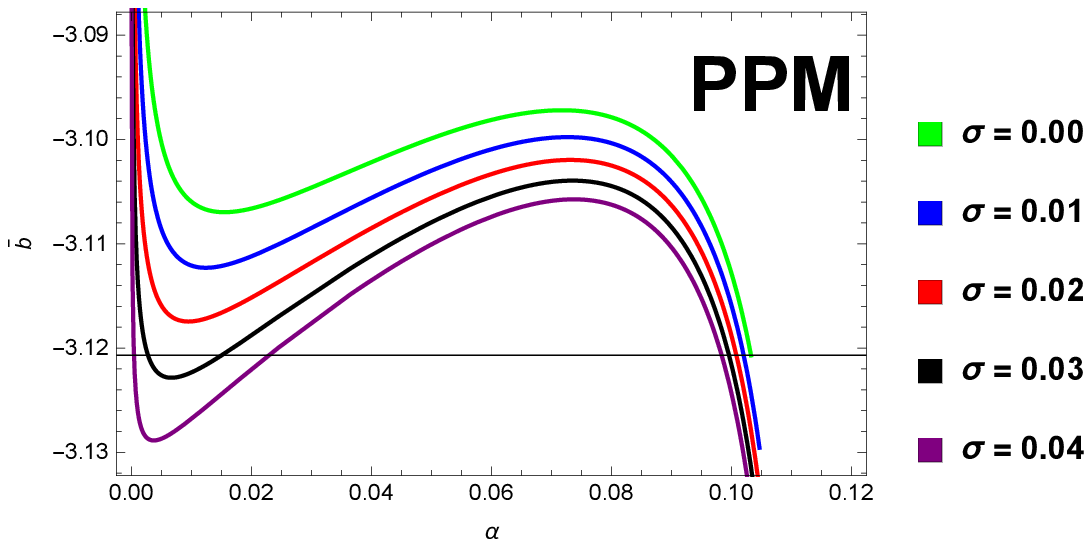}
\includegraphics[width=90mm]{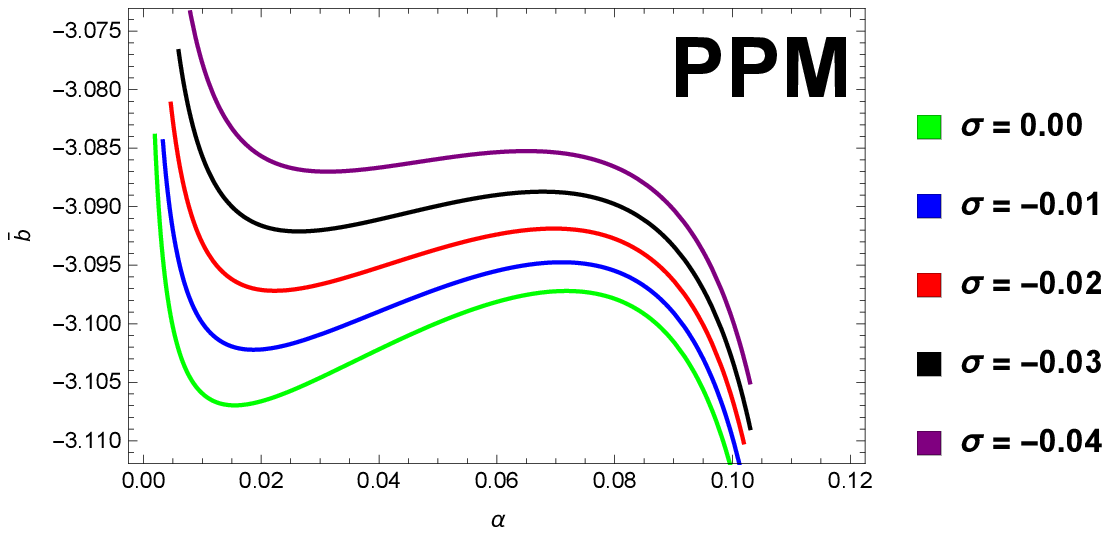}
\includegraphics[width=90mm]{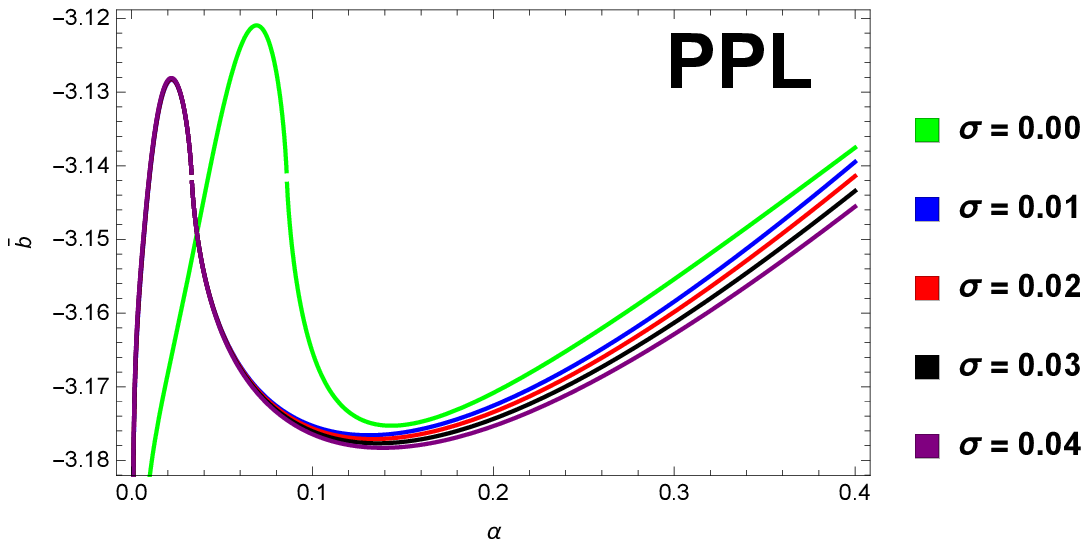}
\includegraphics[width=90mm]{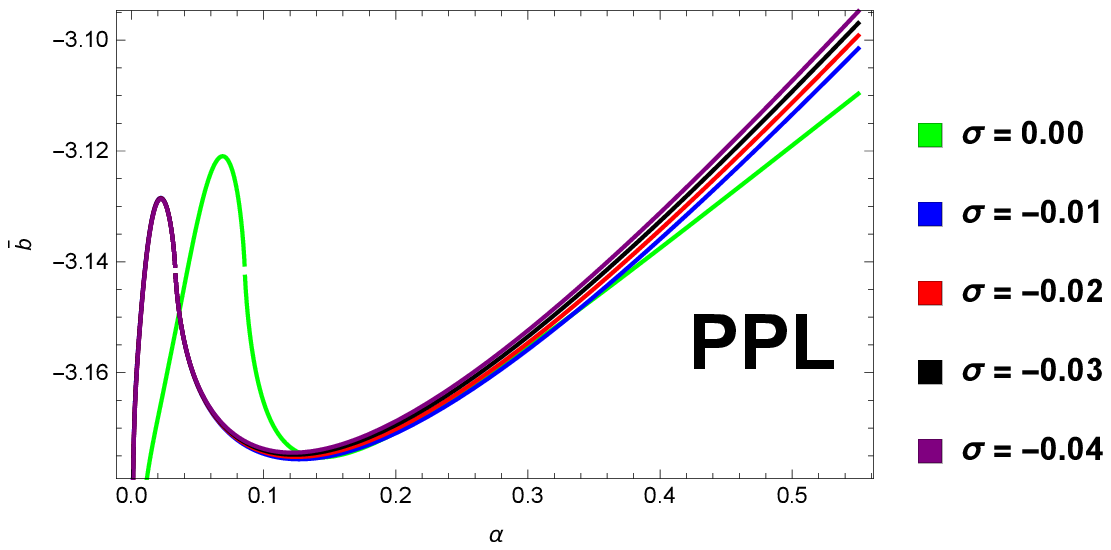}
\caption{Variation of strong deflection limit function $\bar{b}$  with coupling parameter $\alpha$ for different Kiselev parameter $\sigma$
for PPM and PPL cases, where $M = 0.5$.}
\end{figure}
\begin{figure}
\includegraphics[width=90mm]{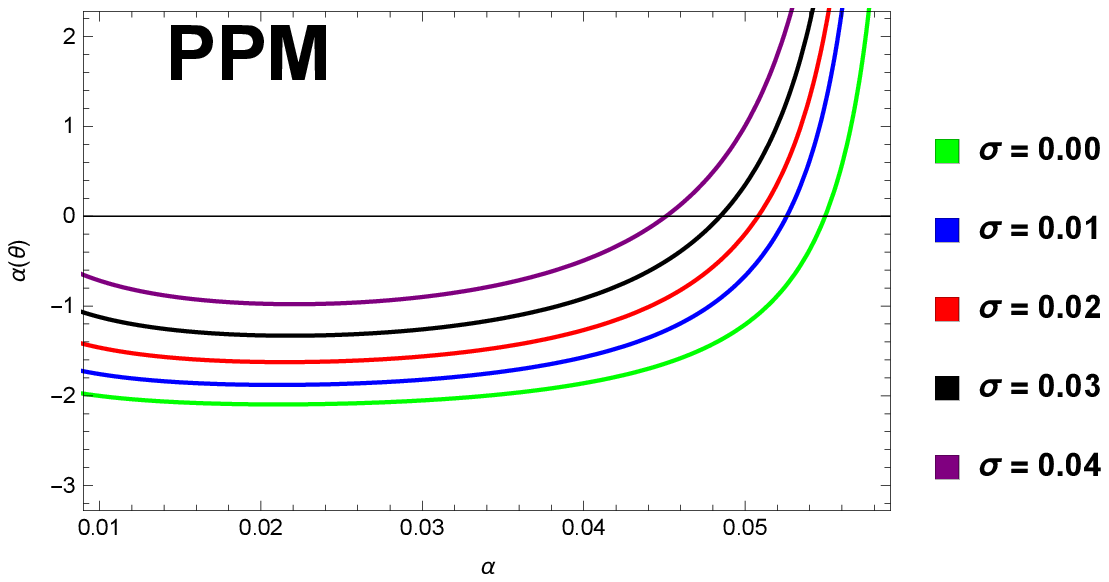}
\includegraphics[width=90mm]{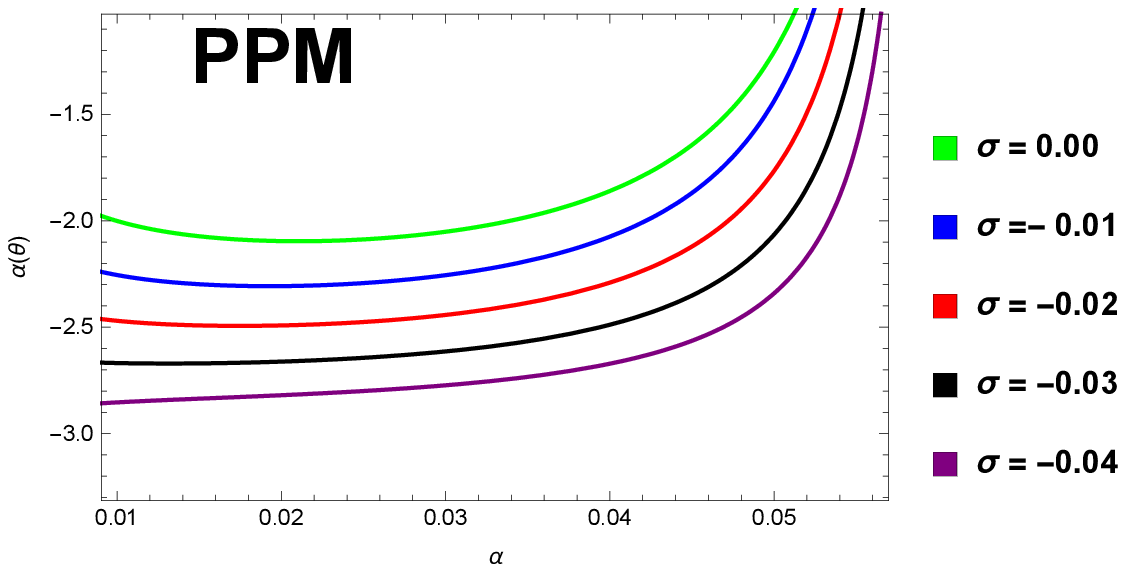}
\includegraphics[width=90mm]{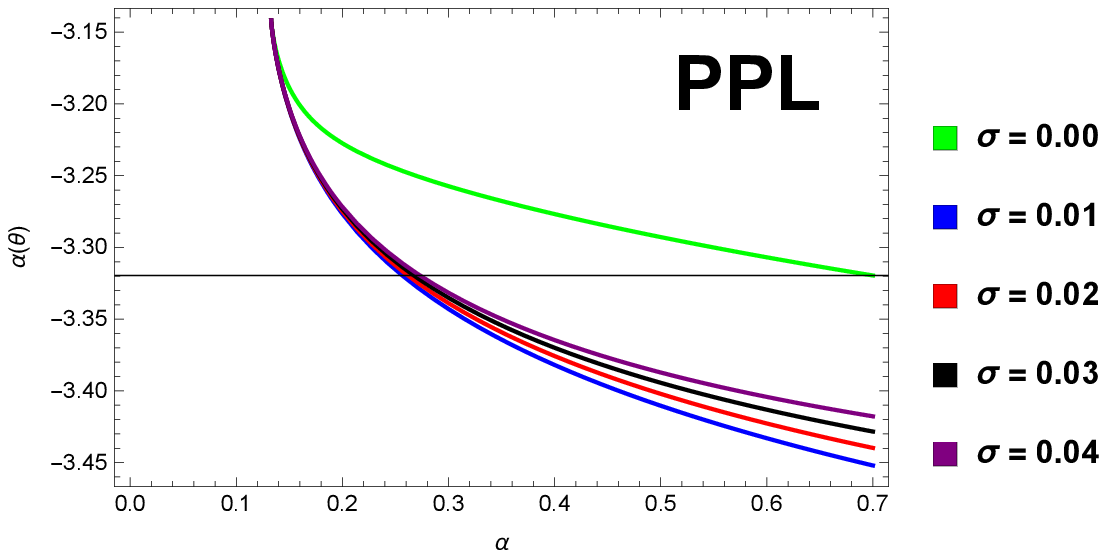}
\includegraphics[width=90mm]{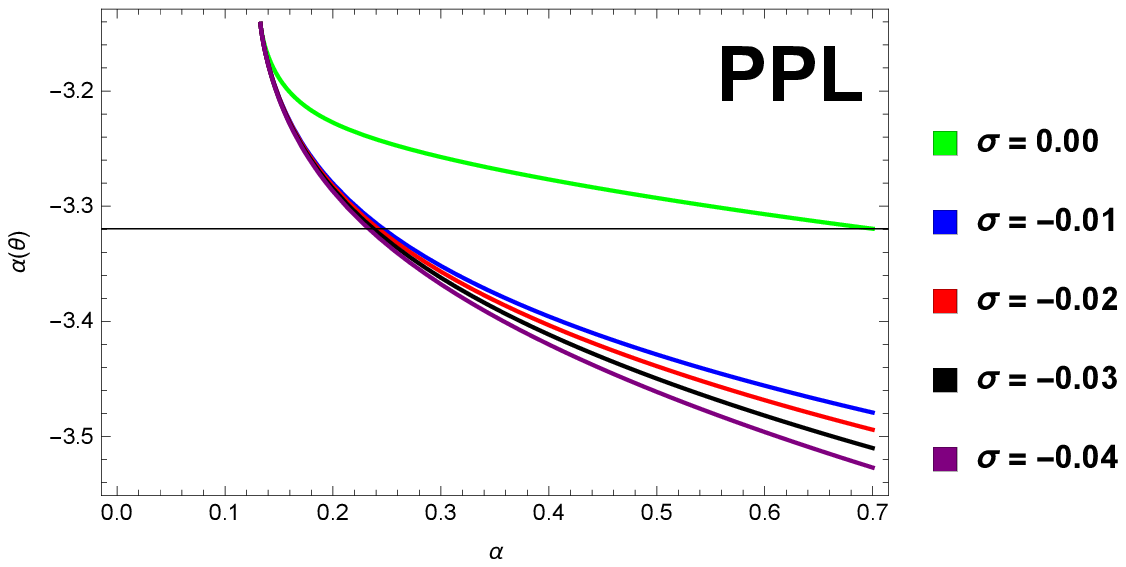}
\caption{Variation of deflection angle $\alpha({\theta})$ with coupling parameter $\alpha$ for different Kiselev parameter $\sigma$
for PPM and PPL cases, where $M = 0.5$.}
\end{figure}
Further, we see that as the coupling parameter $\alpha$ approaches to the critical  values $(i.e., \alpha_{c1}$ or  $\alpha_{c2})$ , the deflection angle can not remain valid in the system $\alpha$ $>\alpha_{c1}$ for the case of PPM and $\alpha$ $<\alpha_{c2}$ for PPL, in the strong deflection limit (67). Hence, with the existence of the coupling terms the variety of the functions $(\bar{a}$ and $\bar{b})$ become more difficult. The reason is that the coupling effects depend not  only on the $\alpha$ values, but also on the coupled photon polarization directions. Moreover, the variation of the deflection angles for PPM and PPL cases is also explored at $u = u_{ps} + 2$, respectively (see Fig.6). We investigate that the behaviors of the deflection angles are same as those for the function $\bar{a}$, which shows that it can be evaluated by the experience that the deflection angles of the photons in the strong field limit are dominated by the logarithmic term.

\subsection{Shadow of Black Hole}
We compute the essential relations to achieve the shape of Kiselev black hole shadow, which suggests the study of motion of the test particle. Further, to obtain the equation of motion, we use the Lagrangian and Hamiltonian Jacobi equation, which demands the study of geodesic equation of the particle near a Kiselev spacetime. So, for describing the motion of particle the Lagrangian $\mathscr{L}$ is given as

\begin{eqnarray}
\mathscr{L} = g_{\mu\nu}\frac{dx^{\mu}}{d\lambda}\frac{dx^{\nu}}{d\lambda}.
\end{eqnarray}
Here, the canonically conjugate momentum for metric (11) have the following form
\begin{eqnarray}
P_{t} = f(r) \dot{t} = E,
\end{eqnarray}
\begin{eqnarray}
P_{r} = f(r)^{-1} \dot{r},
\end{eqnarray}
\begin{eqnarray}
P_{\theta} = r^{2}\dot{\theta} ,
\end{eqnarray}
\begin{eqnarray}
P_{\phi} = r^{2} \sin^{2}\theta \dot{\phi} = L,
\end{eqnarray}
where $E$ known as energy and $L$ defines the angular momentum per unit mass of the photon.
To obtain the circular photon orbits around the particular black hole, the Hamiltonian Jacobi technique is helpful and also, we use the Hamiltonian Jacobi technique to formulate the geodesic equation by applying Carter approach \cite{z5} for Kiselev black hole. In this way, Hamiltonian Jacobi equation can be written for the particular black hole in the following form
\begin{eqnarray}
\frac{\partial \textit{S}}{\partial\lambda} = \mathscr{H} = - \frac{1}{2}g^{\mu\nu}\frac{\partial \textit{S}}{\partial x^{\mu}}\frac{\partial \textit{S}}{\partial x^{\nu}},
\end{eqnarray}
where $\textit{S}$ denotes the action of Jacobi and using Eq.(11) into Eq.(74), we have
\begin{eqnarray}
- 2\frac{\partial \textit{S}}{\partial\lambda} = - \frac{1}{f(r)}(\frac{\partial S_{t}}{\partial t})^2 + f(r ) (\frac{\partial S_{r}}{\partial r})^2 + \frac{1}{r^2} (\frac{\partial S_{\theta}}{\partial \theta})^2 + \frac{1}{r^2\sin^2 \theta}(\frac{\partial S_{\phi}}{\partial \phi})^2.
\end{eqnarray}
Now, we suppose two Killing fields $\xi_{t,\phi} = \partial_{t,\phi}$ for simplicity, then the action of Jacobi $\textit{S}$ takes the form
\begin{eqnarray}
\textit{S} = \frac{1}{2}m^{2}_{p} \lambda - E t + \textit{S}_{r}(r) + \textit{S}_{\theta}(\theta) + L \phi,
\end{eqnarray}
\begin{figure}
\includegraphics[width=58mm]{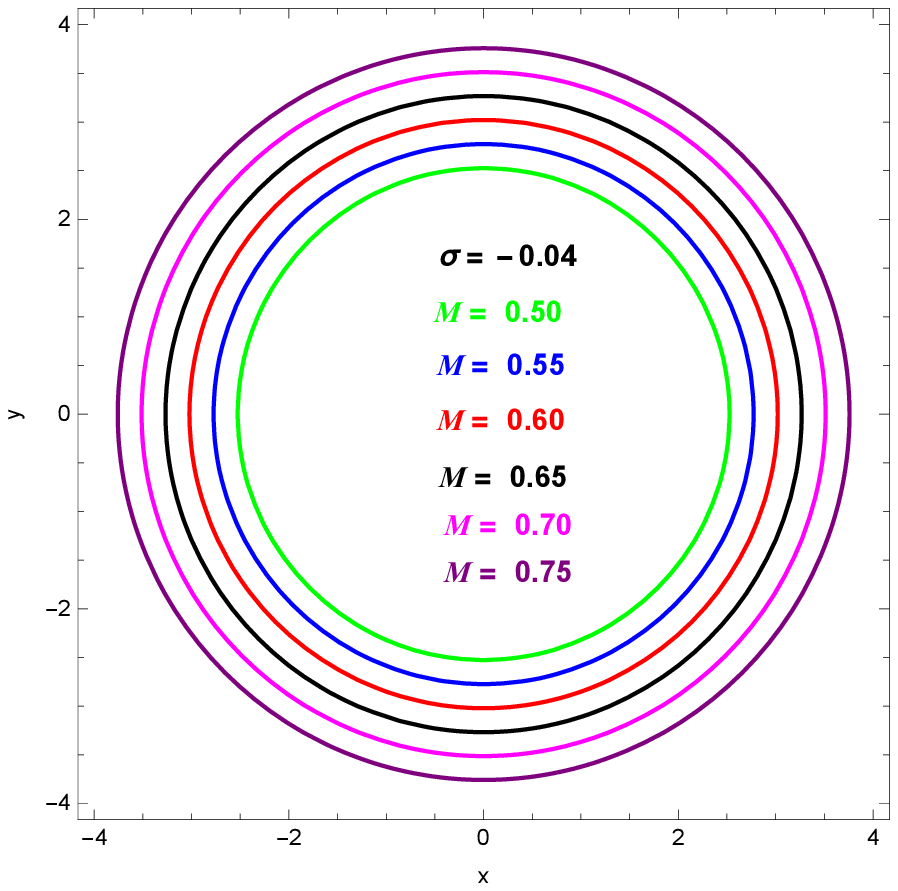}
\includegraphics[width=58mm]{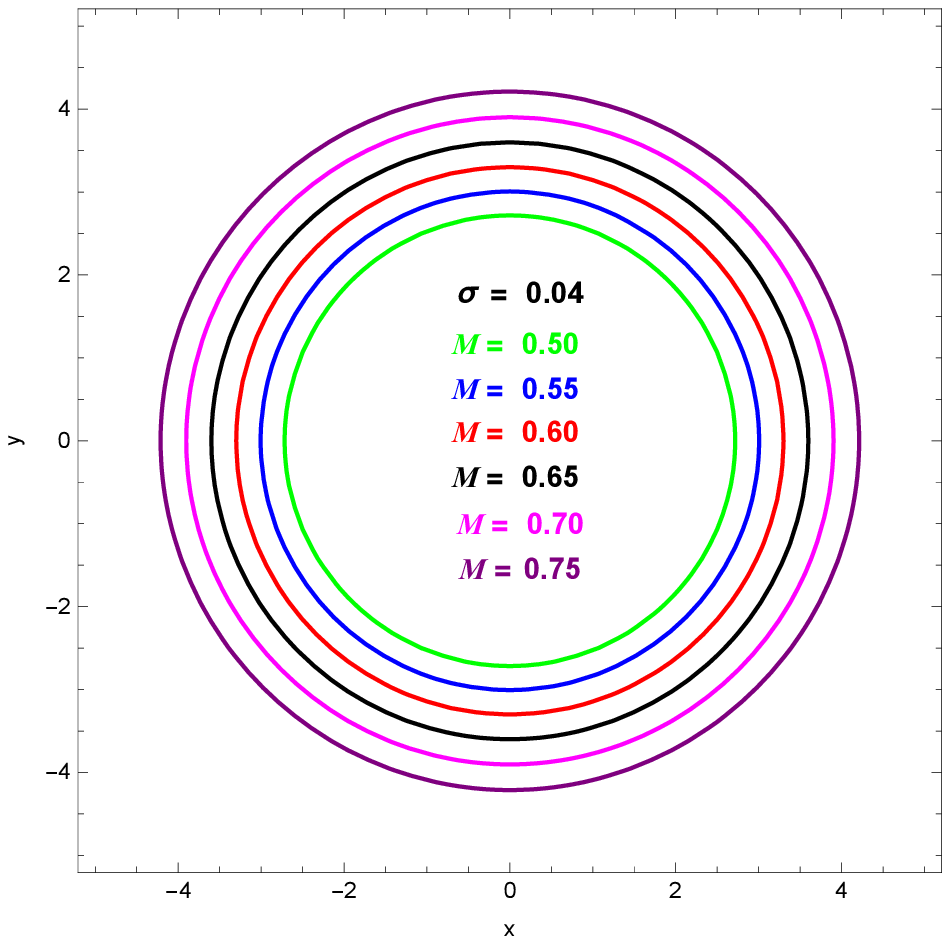}
\includegraphics[width=58mm]{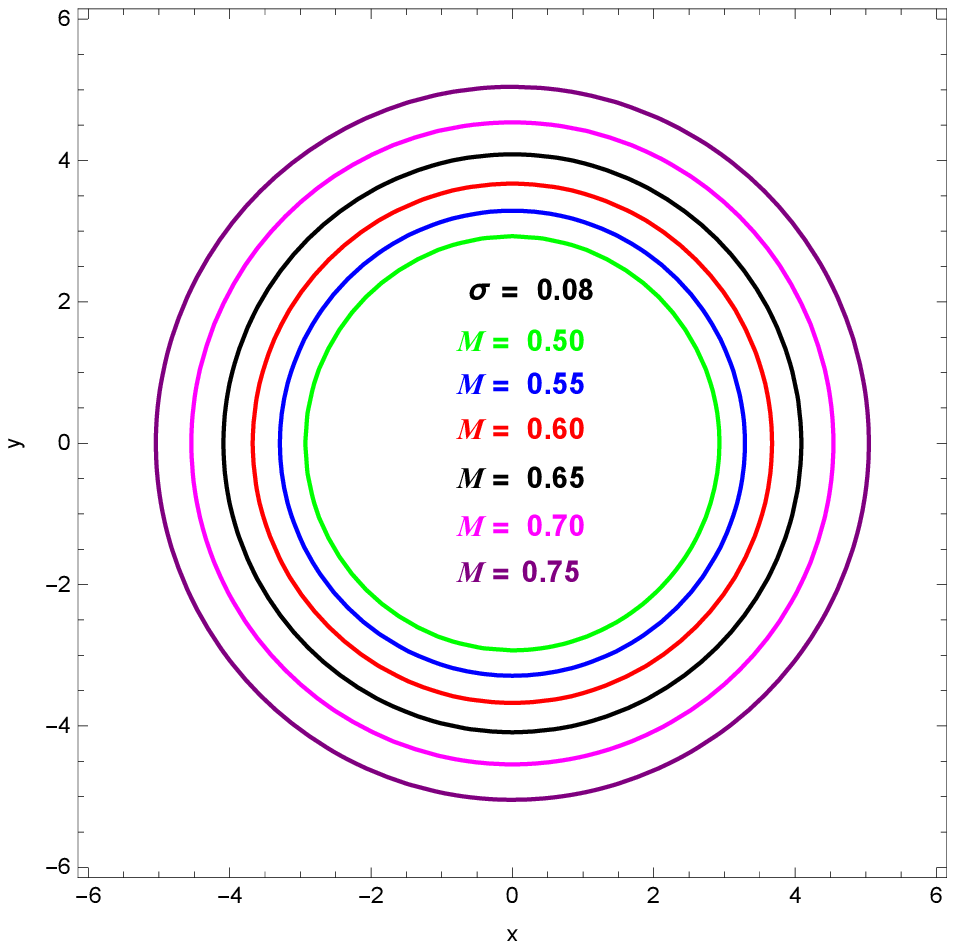}
\includegraphics[width=58mm]{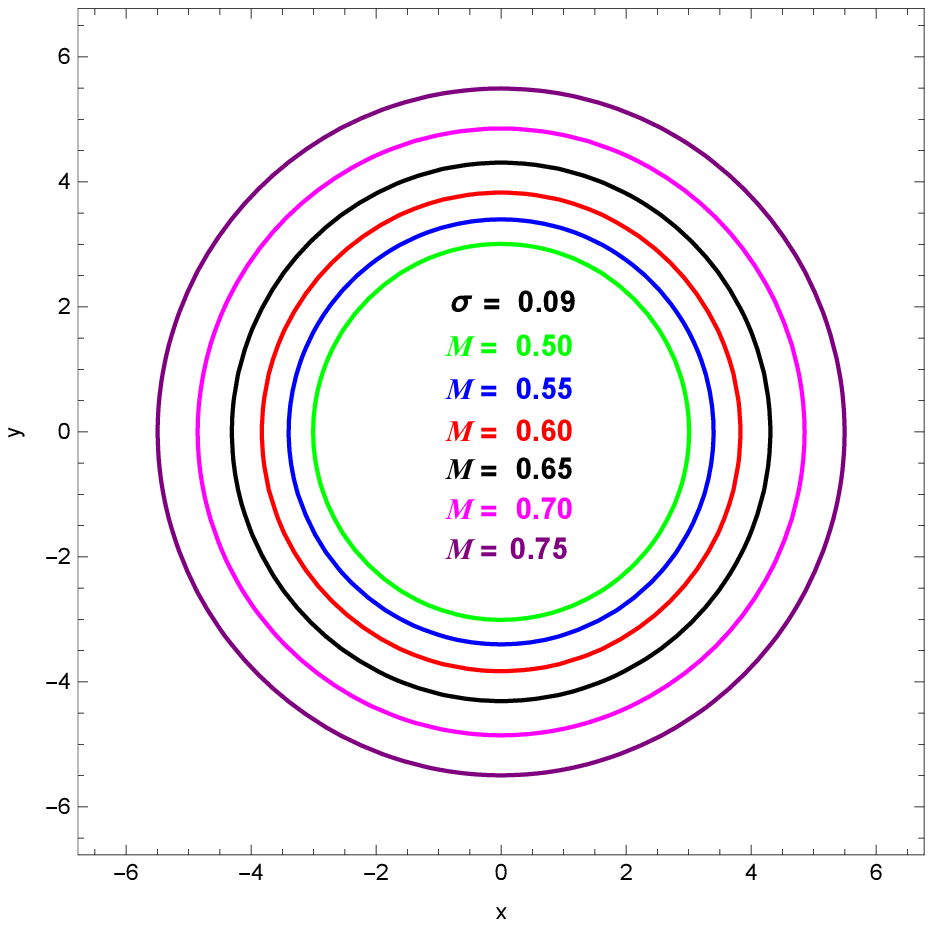}
\includegraphics[width=58mm]{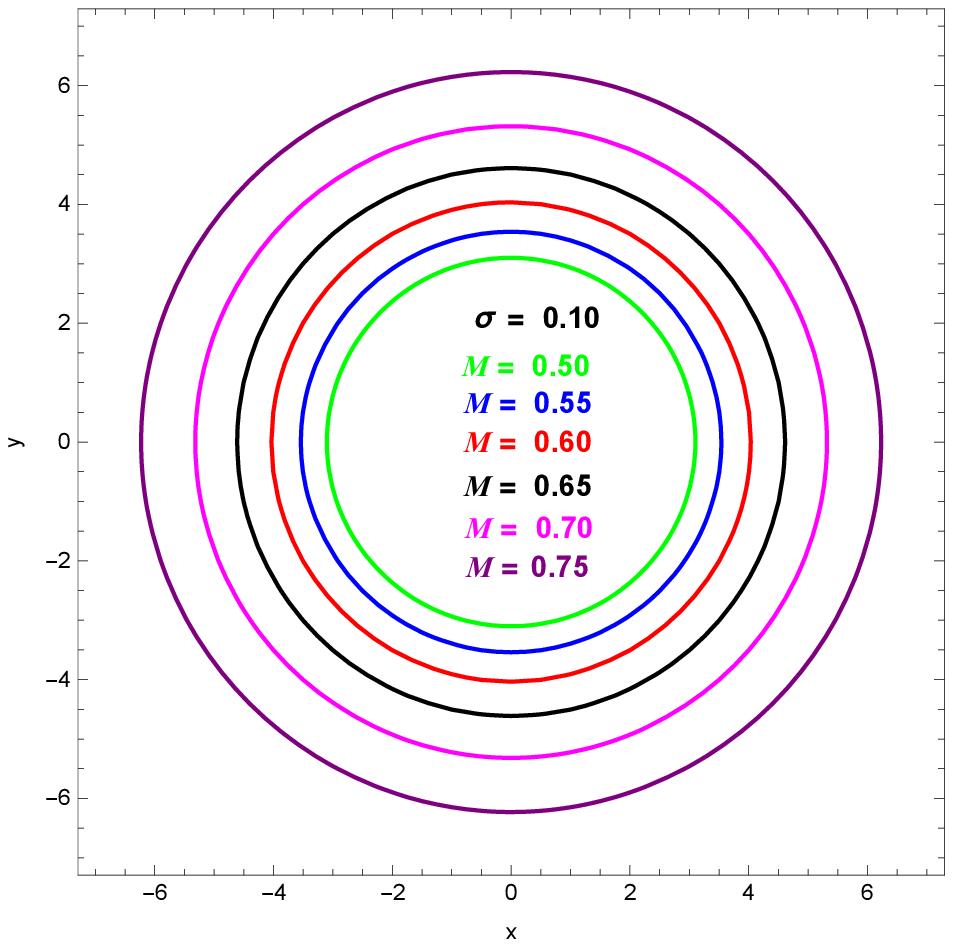}
\includegraphics[width=58mm]{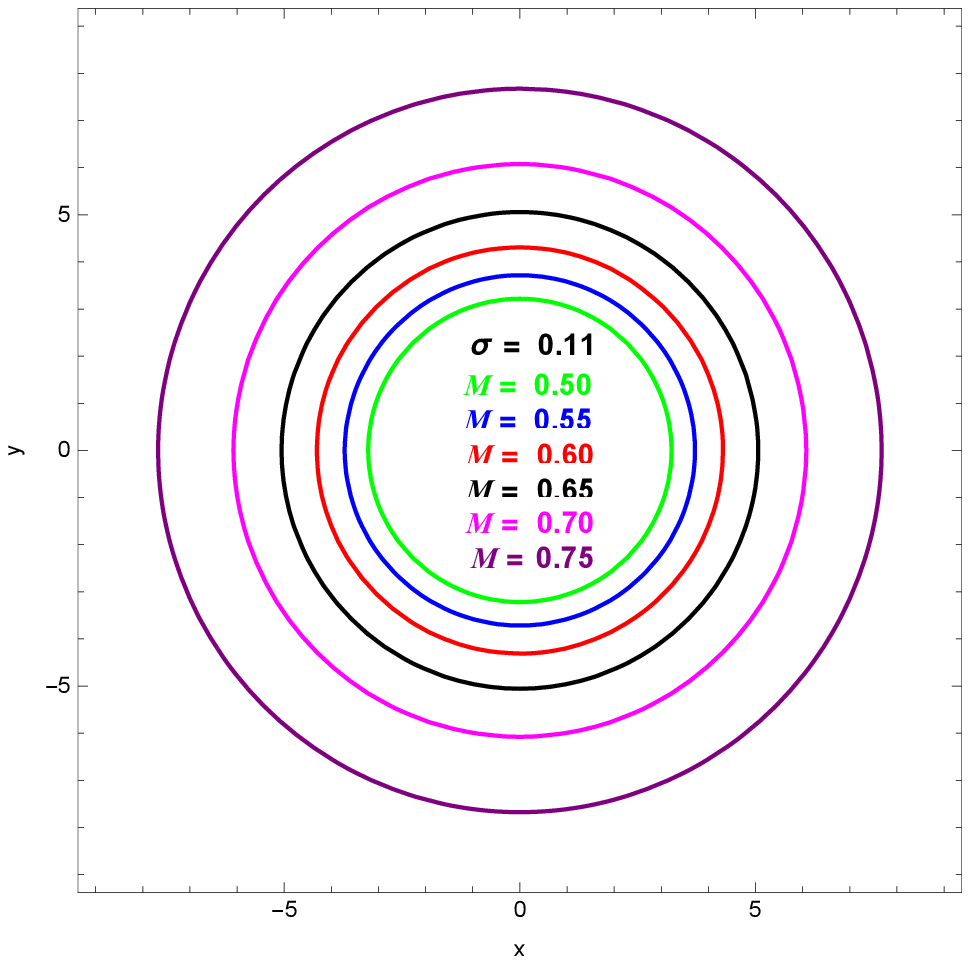}
\includegraphics[width=58mm]{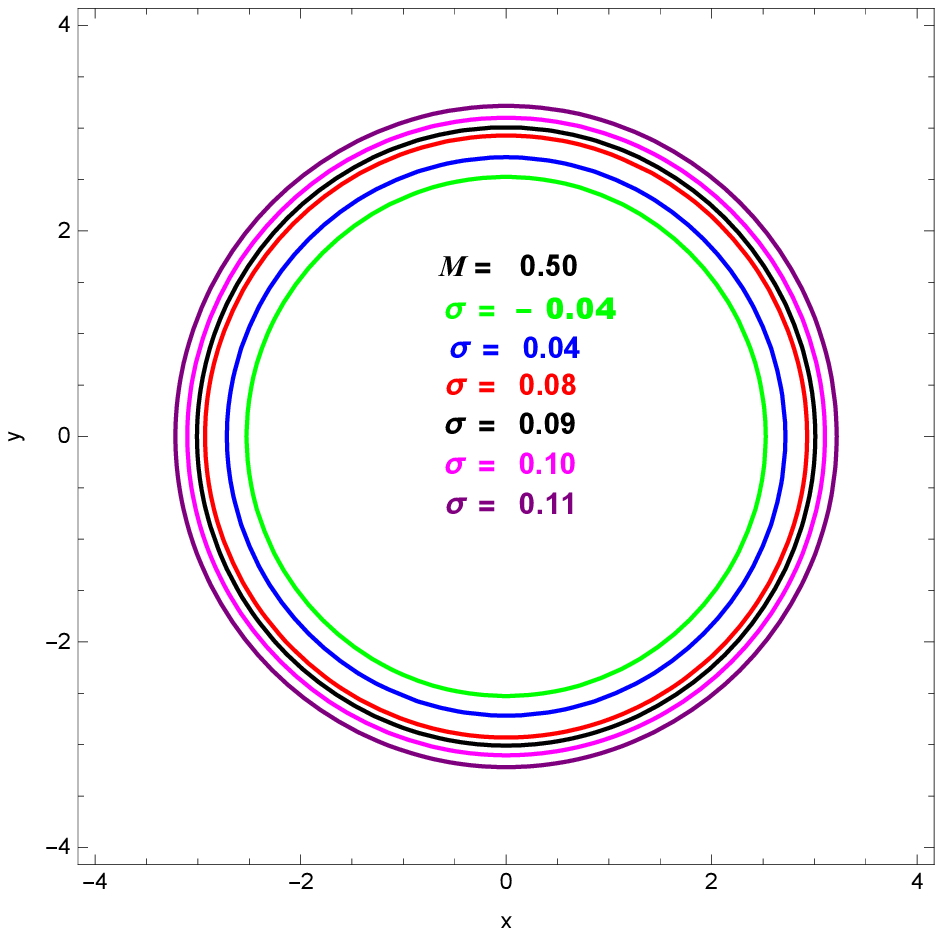}
\includegraphics[width=58mm]{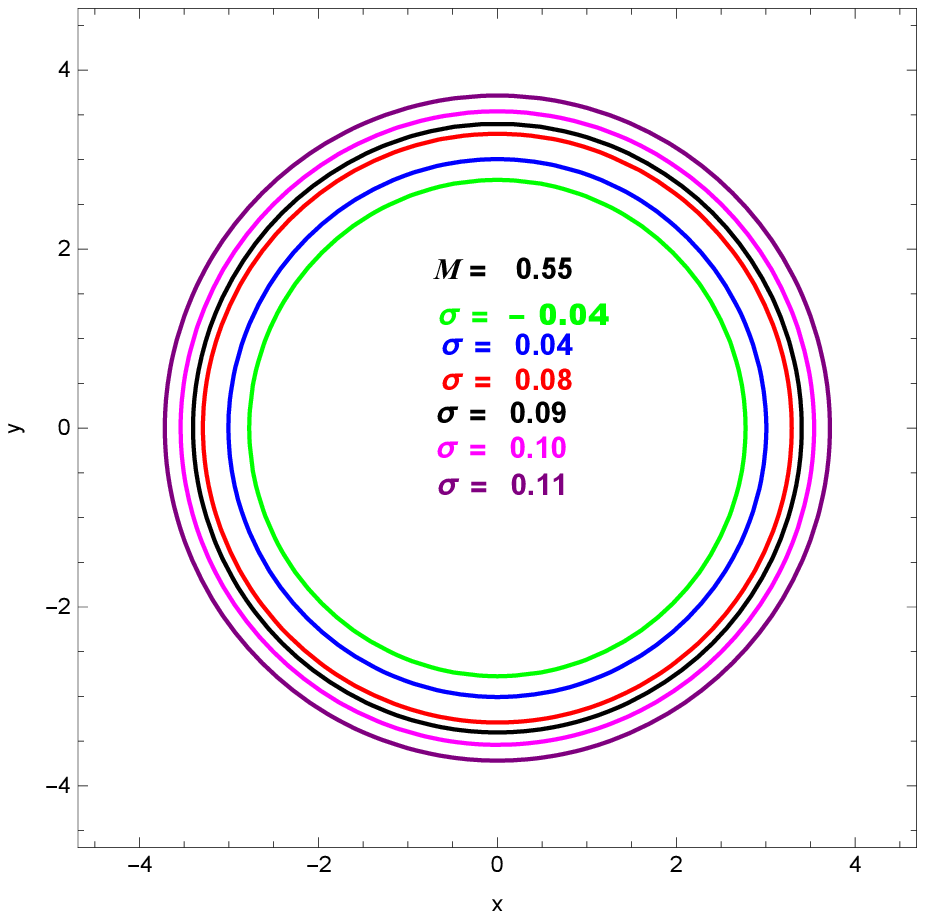}
\includegraphics[width=58mm]{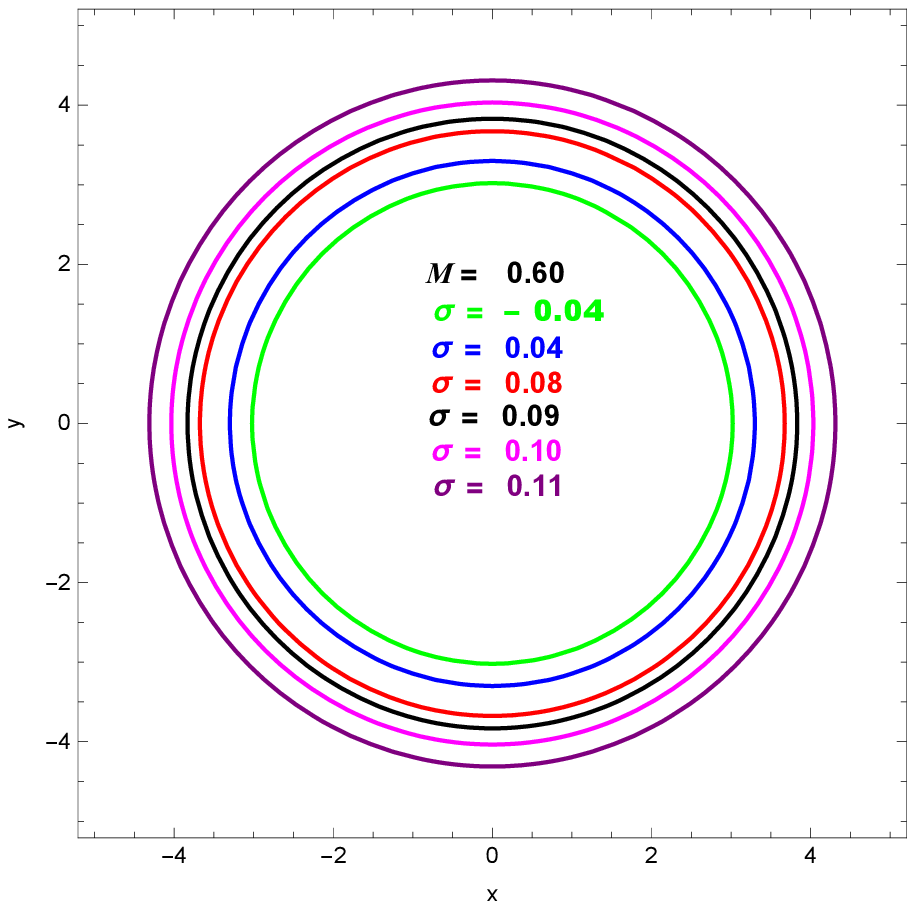}
\includegraphics[width=58mm]{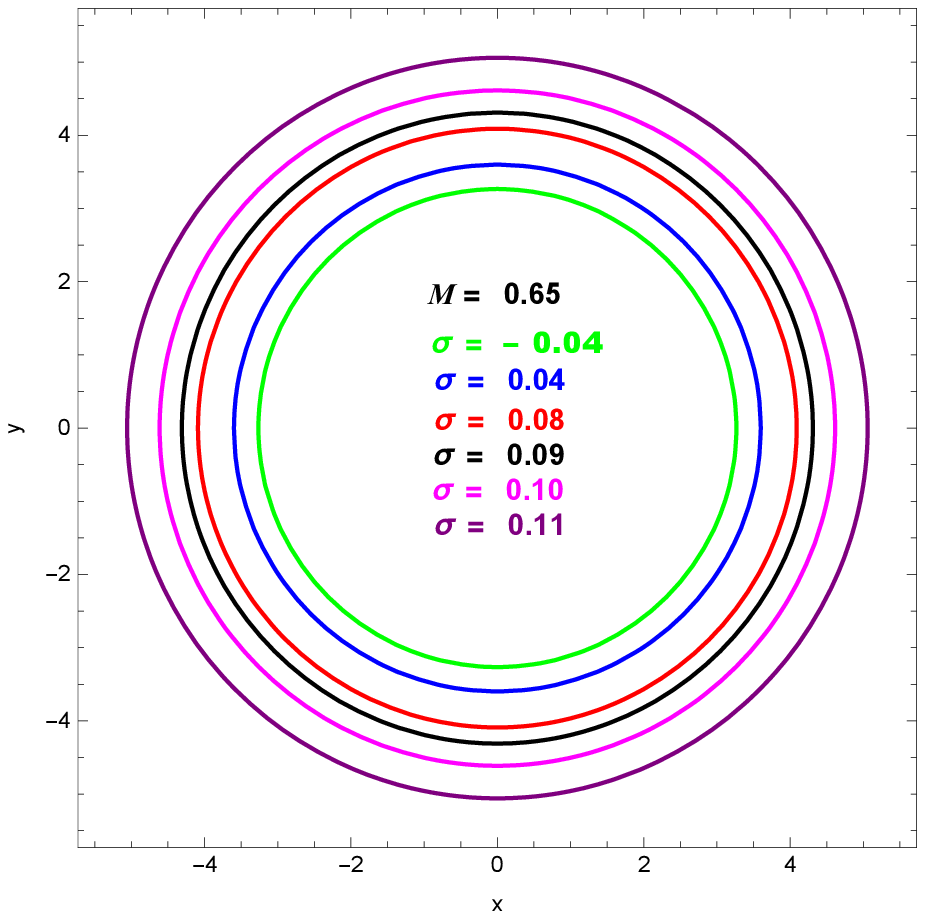}
\includegraphics[width=58mm]{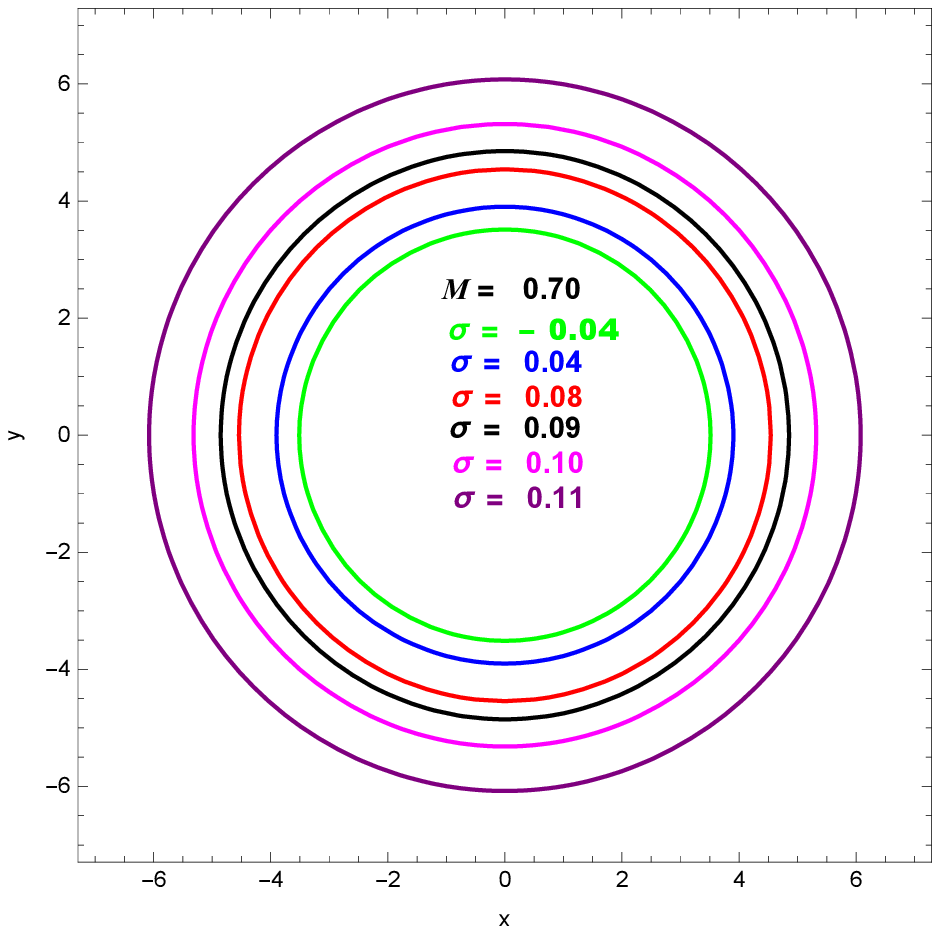}
\includegraphics[width=58mm]{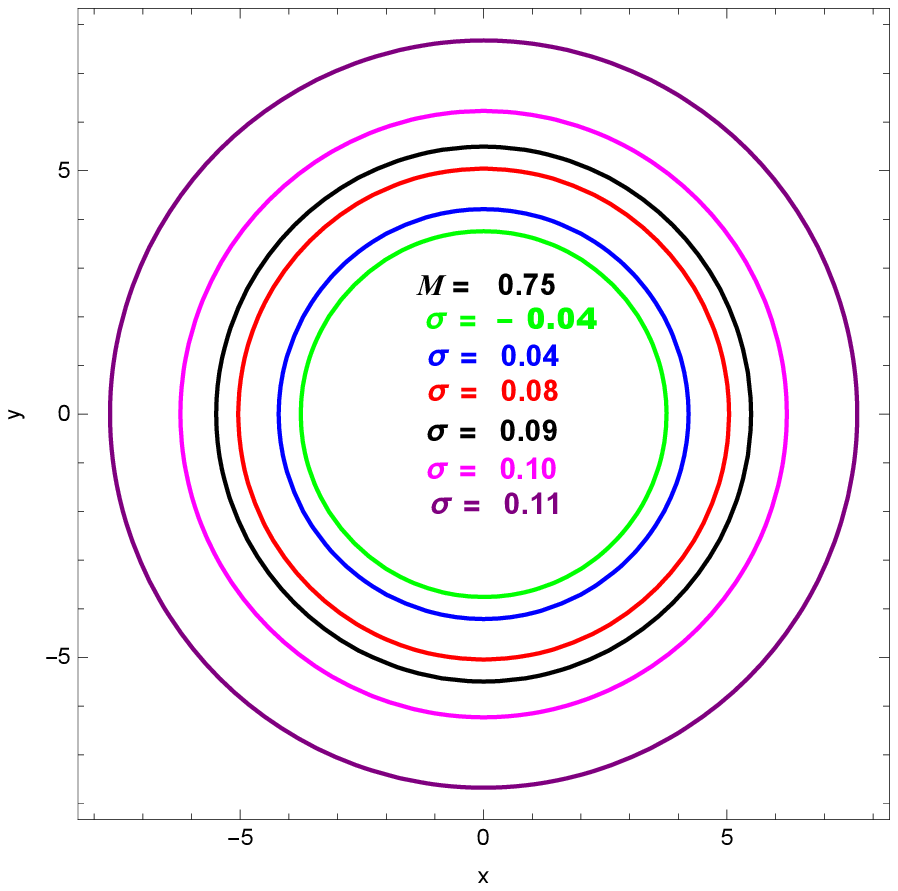}
\caption{Shadow cast by Kiselev black hole spacetime at $\theta = \pi/2$, for different values of the Kiselev parameter $\sigma$ and mass $M$. }
\end{figure}
where $m_{p}$ is a test particle mass, for the photon it is assumed to be zero and $\textit{S}_{r}(r)$, $\textit{S}_{\theta}(\theta)$ both are the functions of $r$ and $\theta$, respectively. Inserting the Eq.(76) into Eq.(74), we get the complete equation of null geodesic for Kiselev black hole spacetime
\begin{eqnarray}
\dot{t} = \frac{E}{f(r)},
\end{eqnarray}
\begin{eqnarray}
r^2 \dot{r} = \pm\sqrt{R},
\end{eqnarray}
\begin{eqnarray}
r^2 \dot{\theta} = \pm\sqrt{\Theta},
\end{eqnarray}
\begin{eqnarray}
\dot{\phi} = \frac{L}{r^2\sin^2 \theta},
\end{eqnarray}
where the signs $\pm$ are used for the radial direction of outgoing and ingoing particle motion, respectively \cite{z6}. Here, $R$ and $\Theta$ are defined by
\begin{eqnarray}
R = r^4 E^{2} -  r^{2}(1 - \frac{2M}{r} - \sigma r)(L^{2} + \mathcal {K})
\end{eqnarray}
\begin{eqnarray}
\Theta = \mathcal {K} -  L^{2} \cot^{2} \theta,
\end{eqnarray}
where $\mathcal {K}$ is called the Carter constant. The effective potential $V_{eff}(r)$ for the asymptotically non-flat spacetime (11), is defined as follows:
\begin{eqnarray}
V_{eff}(r) = \frac{1}{r^{2}}(1 - \frac{2M}{r} - \sigma r)(L^{2} + \mathcal {K}) - E^{2}.
\end{eqnarray}
The unstable circular orbits constitute the photon sphere and define the boundary of the shadow cast by the compact body. These unstable circular orbits can be obtained by maximizing the effective potential $V_{eff}(r)$, which leads to
\begin{eqnarray}
V_{eff}(r) = \frac{\partial V_{eff}(r)}{\partial r} = 0  &or&  R(r) = \frac{\partial R(r)}{\partial r} = 0.
\end{eqnarray}

For the general orbits, we consider two impact parameters $\xi = \frac{L}{E}$ and $\eta = \frac{\mathcal {K}}{E^{2}}$, which are the functions of the constants of motion $E$, $L$ and $\mathcal {K}$. These impact parameters define the properties of the photons near the black hole. To visualize the black hole shadow more clearly, it is useful to adopt the celestial coordinates. Recently, Haroon et al. \cite{z7} have introduced the technique for defining the celestial coordinates of asymptotically non-flat spacetime, we follow this approach and find the following form of modified celestial coordinates
\begin{eqnarray}
x = -\sqrt{1-\sigma}\xi csc\theta,
\end{eqnarray}
\begin{eqnarray}
y = \pm\sqrt{(1-\sigma}\sqrt{\eta - \xi^{2}\cot^{2}\theta},
\end{eqnarray}
For the equatorial plane $\theta = \frac{\pi}{2}$, the Eqs.(85) and (86) are simplified as
\begin{eqnarray}
x = -\sqrt{1-\sigma}\xi,
\end{eqnarray}
\begin{eqnarray}
y = \pm\sqrt{(1-\sigma}\sqrt{\eta}.
\end{eqnarray}
The Eqs. (87) and (88) yield the following relation
\begin{eqnarray}
x^{2} + y^{2} =(1-\sigma M) \frac{(-2-5\sigma M+2\sqrt{1+6\sigma M})(-1+\sqrt{1+6\sigma M})^{3}}{\sigma^{2}(-1-2\sigma M+\sqrt{1+6\sigma M})(-3-8\sigma M+3\sqrt{1+6\sigma M})}.
\end{eqnarray}

The contour of the Eq. (89) can describe the apparent shape of Kiselev
black hole. From Eq. (89), the size of Kiselev black hole depends upon mass and Kiselev parameter
of space-time. The Eq. (89) governs the complete orbit of photon around black hole which cast shadow
and appears as circle. Now we take the contour plot of Eq. (89) which shows the shadow of
Kiselev black hole, clearly shown in Fig. 7.  The size of the shadow cast by Kiselev black hole increases with the increase of Kiselev parameter and mass.
\subsection{Magnification Factors}
The lens equation for the observer and the source can be written in the following form \cite{28}
\begin{eqnarray}
\gamma = \frac{D_{1} + D_{2}}{D_{2}}\theta - \alpha(\theta) mod 2\pi,
\end{eqnarray}
where $\gamma$ defines the angle between the optical axis and the source direction, $D_{1}$ represents the distance of observer and lens, $D_{2}$ represents the distance of the source and lens. Here, we take only the simplest case in which the observer, lens and source are extremely arranged, so that the angular separation for $n^{th}$ relativistic image and the lens may be defined as \cite{28}
\begin{figure}
\includegraphics[width=90mm]{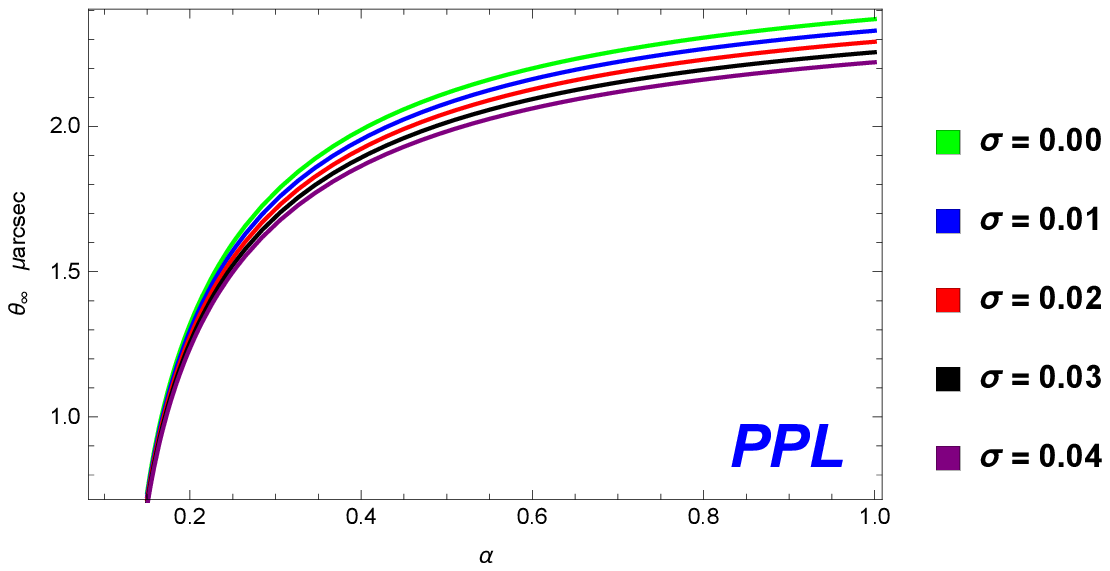}
\includegraphics[width=90mm]{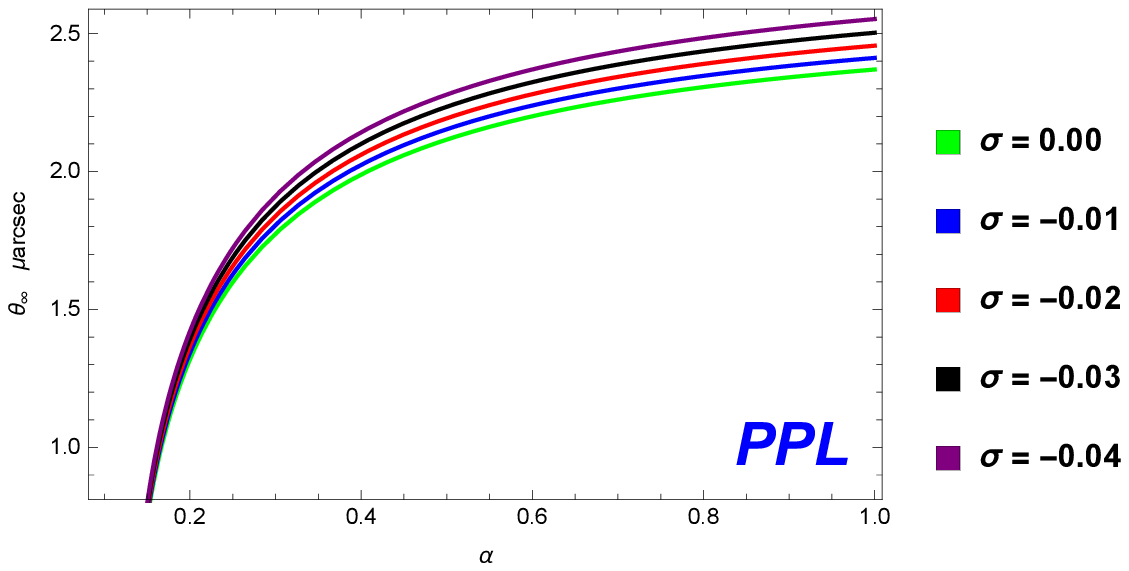}
\includegraphics[width=90mm]{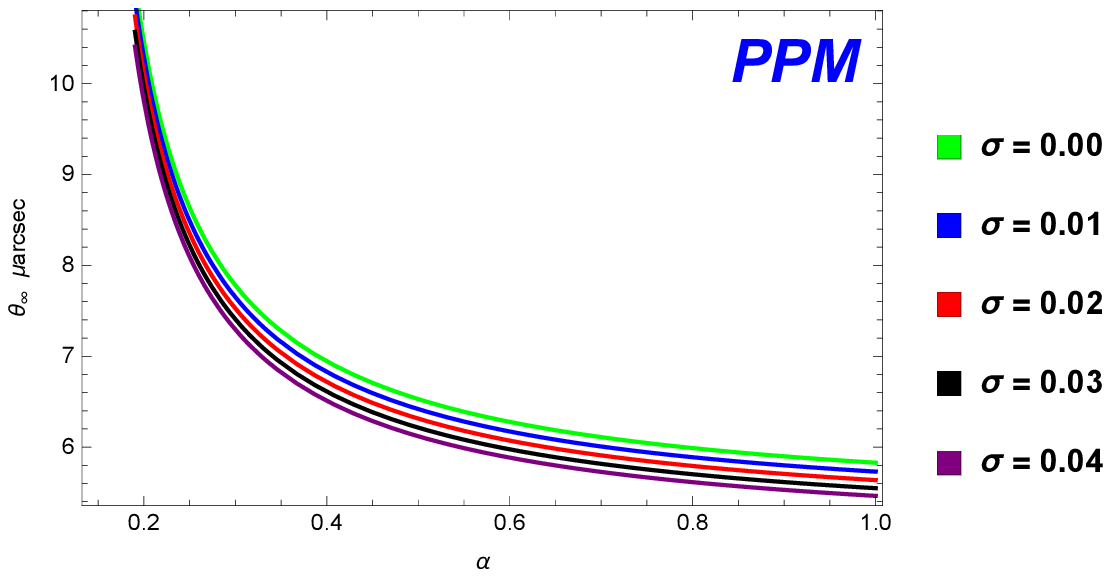}
\includegraphics[width=90mm]{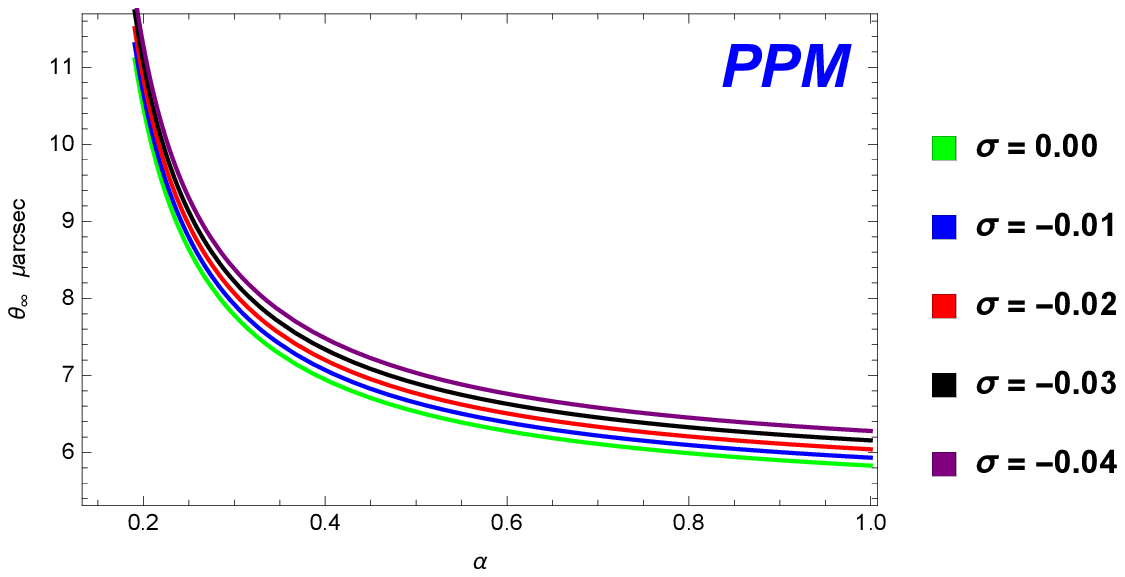}
\caption{Variation of innermost relativistic image $\theta_{\infty}$  with coupling parameter $\alpha$ for different Kiselev $\sigma$ for the cases of PPL and PPM, where $M = 0.5$.}
\end{figure}
\begin{figure}
\includegraphics[width=90mm]{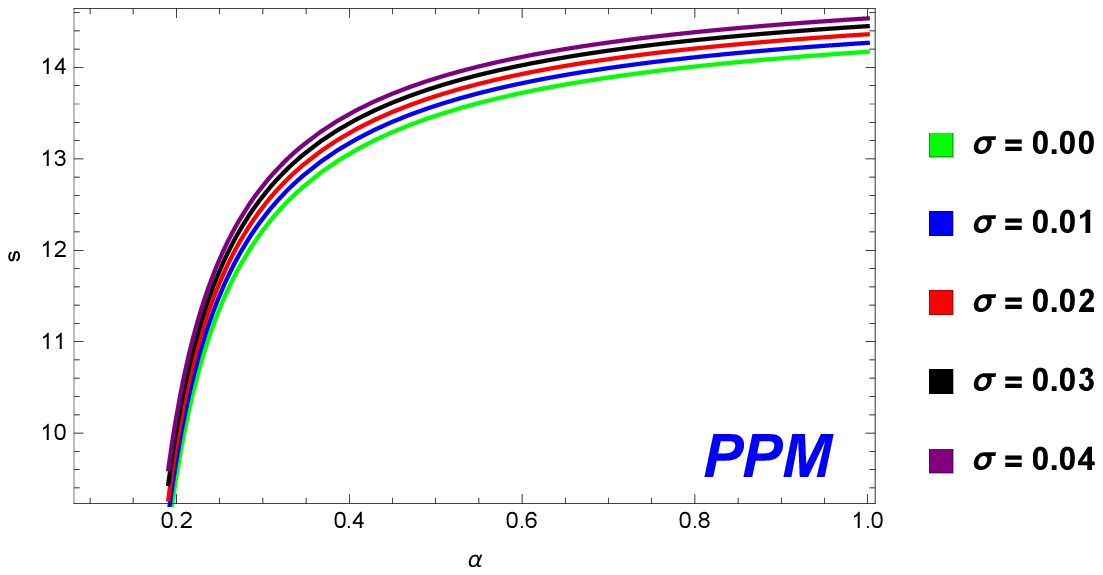}
\includegraphics[width=90mm]{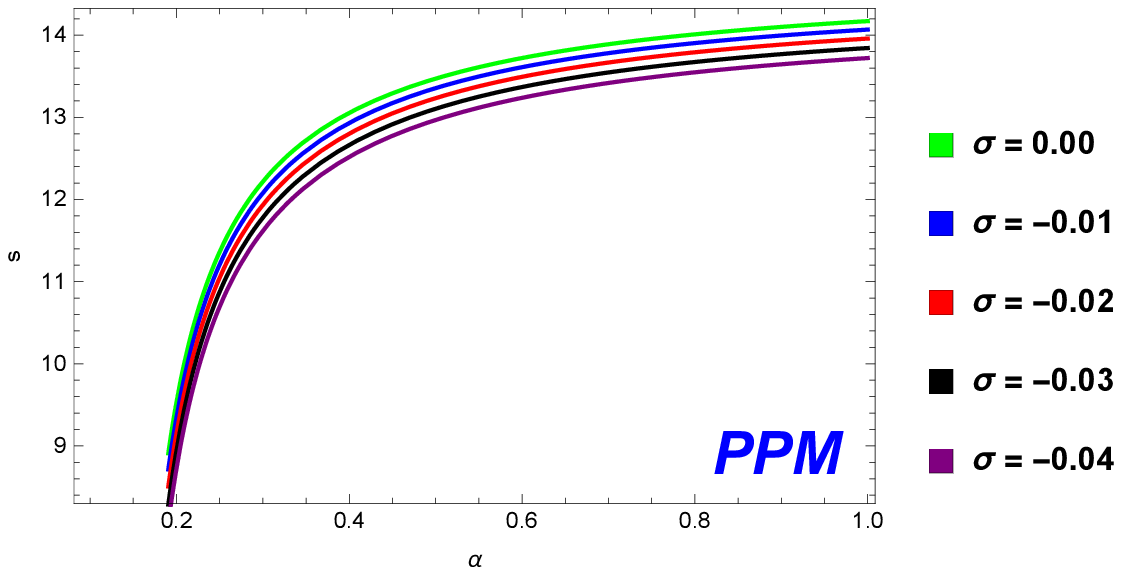}
\includegraphics[width=90mm]{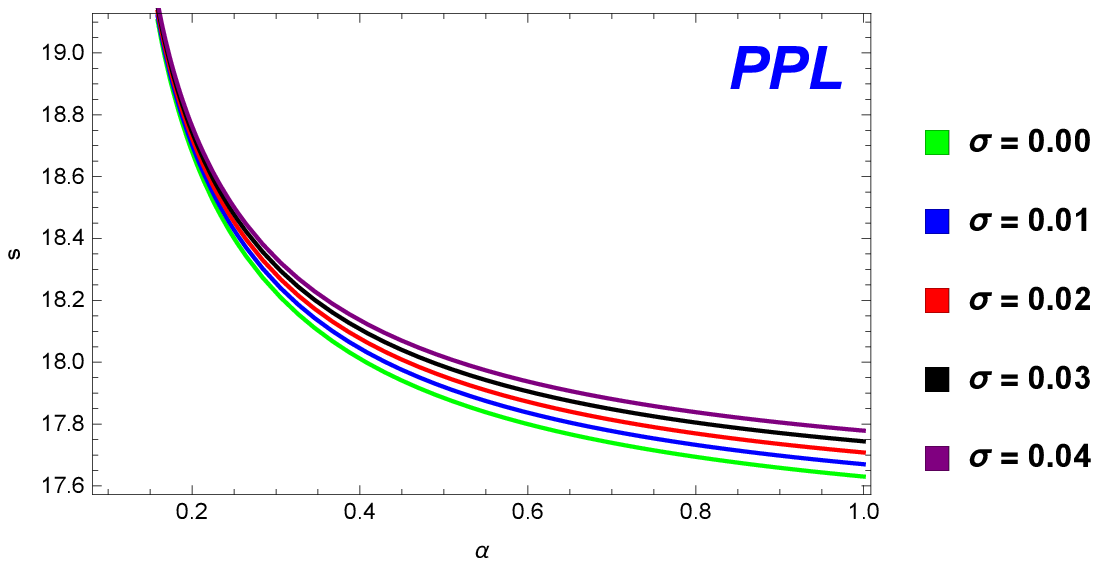}
\includegraphics[width=90mm]{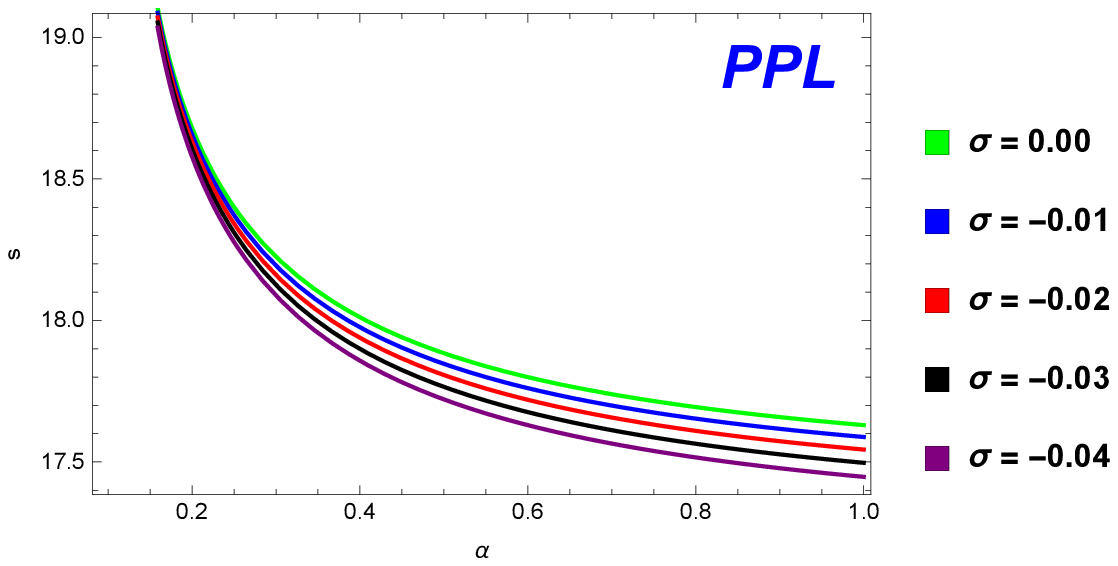}
\caption{Variation of angular separation s  with coupling parameter $\alpha$ for different Kiselev $\sigma$ for the cases of PPM and PPL, where $M = 0.5$.}
\end{figure}
\begin{figure}
\includegraphics[width=88mm]{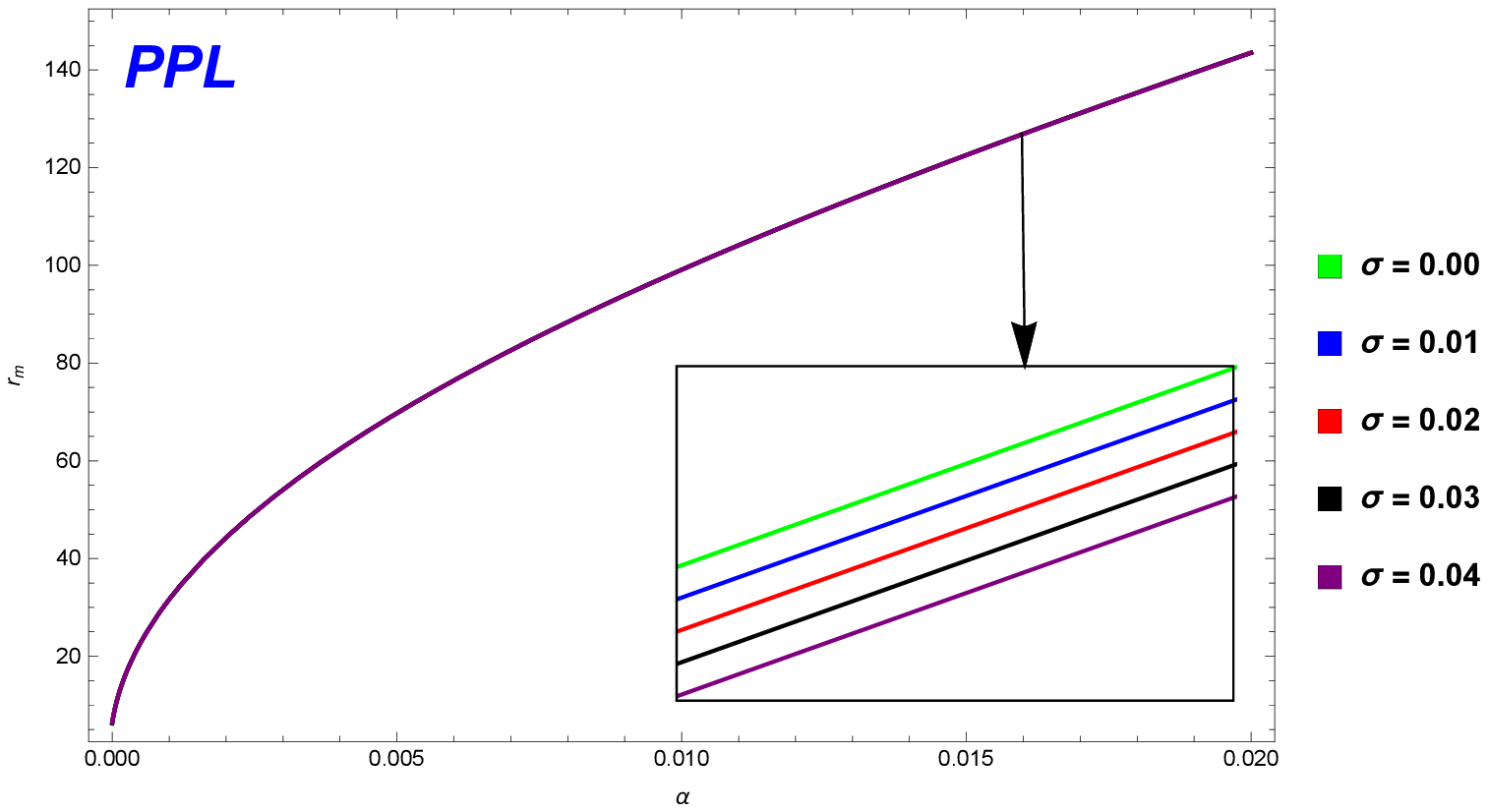}
\includegraphics[width=90mm]{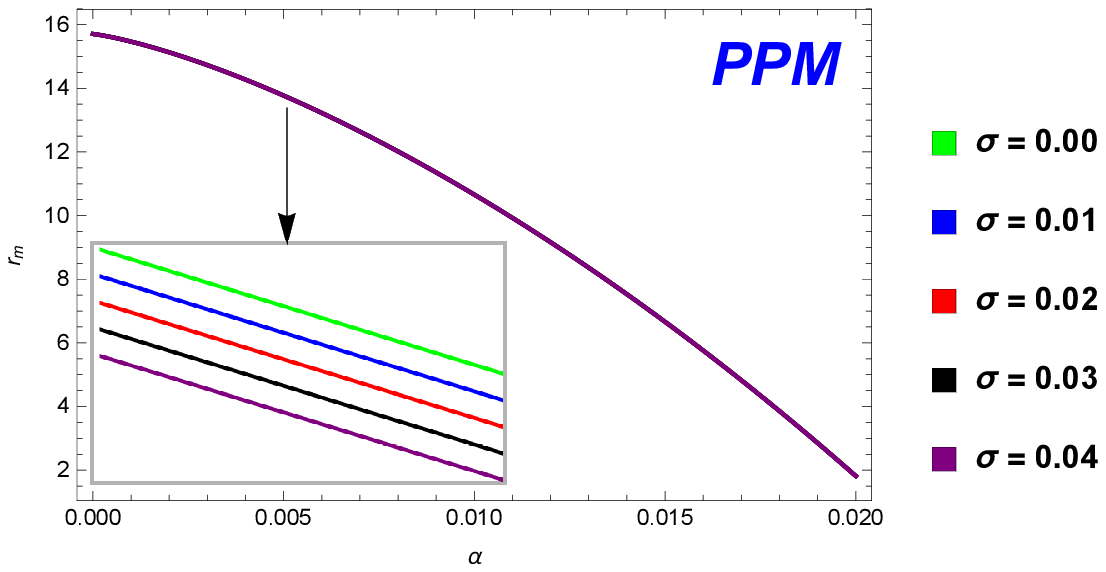}
\includegraphics[width=93mm]{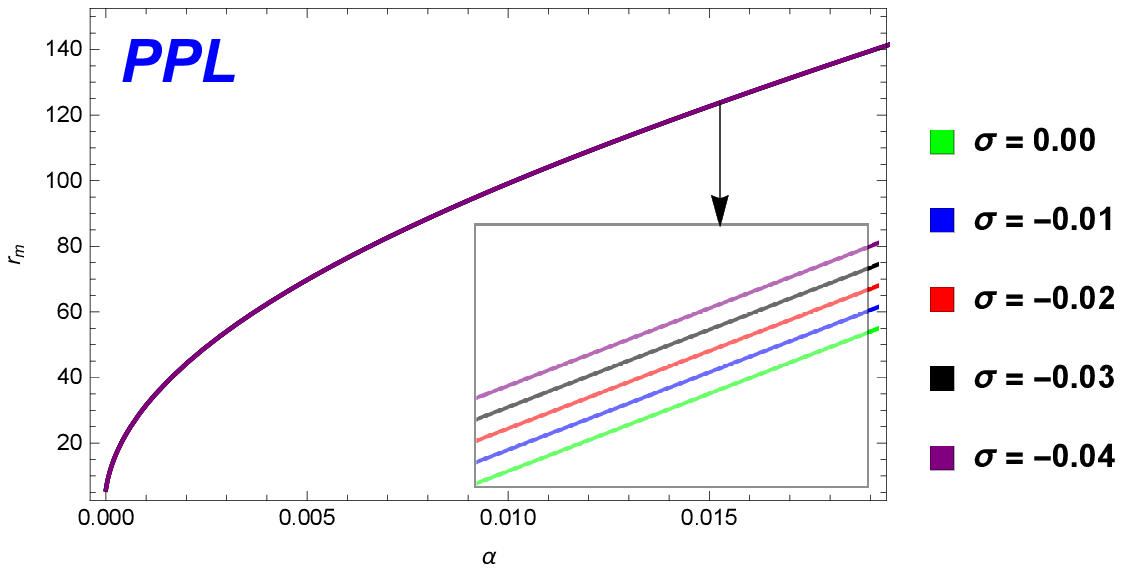}
\includegraphics[width=90mm]{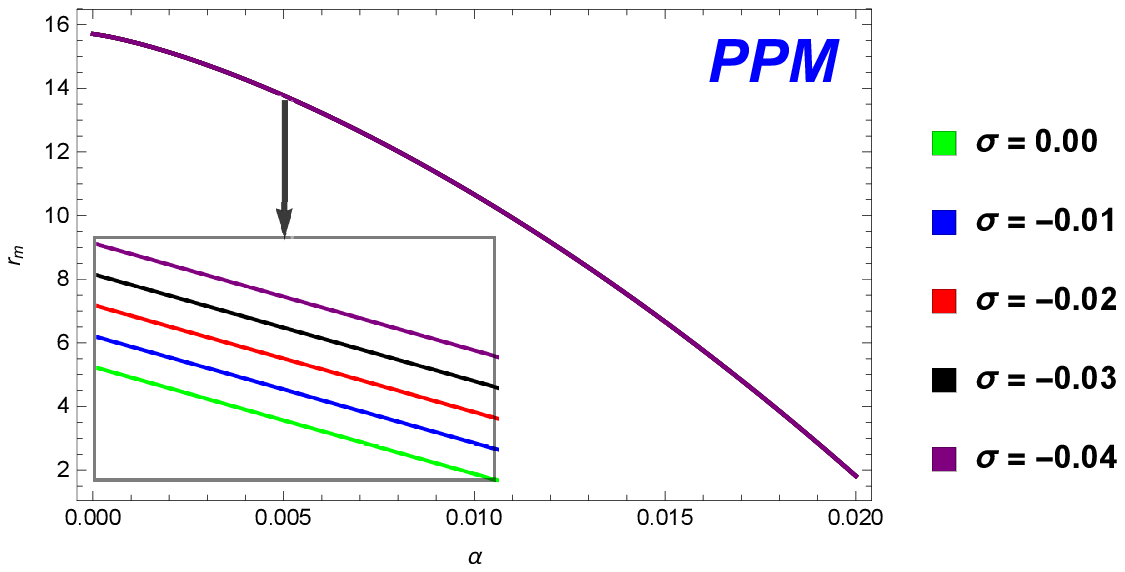}
\caption{Variation of relative magnitude $r_{m}$  with coupling parameter $\alpha$ for different Kiselev parameter $\sigma$ for the cases of PPL and PPM, where $M = 0.5$.}
\end{figure}
\begin{eqnarray}
\theta_{n} \simeq \theta^{0}_{n}(1 - \frac{u_{ps}e_{n}(D_{1} + D_{2}) }{\bar{a}D_{1}D_{2}}),
\end{eqnarray}
and
\begin{eqnarray}
\theta^{0}_{n} = \frac{u_{ps}}{D_{1}}(1 + e_{n}), && e_{n} = e^{\frac{\bar{b} + |\gamma| - 2 \pi n}{\bar{a}}},
\end{eqnarray}
where the position of image alternate to $\alpha = 2n\pi$ is $\theta^{0}_{n}$ and $n$ is any integer.  If $n\rightarrow \infty$, we
obtain the relation $e_{n}\rightarrow 0$. This relation gives a result for the impact parameter $u_{ps}$, distance $D_{1}$ and a set of images $\theta_{\infty}$, which can be defined as
\begin{eqnarray}
u_{ps} = D_{1}\theta_{\infty}.
\end{eqnarray}
We investigate that the strong deflection limit functions $\bar{a}$ and $\bar{b}$, which can be obtained if there exist extra two observations. Thus as in \cite{27,28}, we
suppose a perfect situation where the outermost image $\theta_{1}$ is separated as a single image and all
the remaining ones are packed together at relativistic images $\theta_{\infty}$. In this way, the angular separation $s$ and the relative
magnitudes $r_{m}$ among the first image and other ones may be further defined as
\begin{eqnarray}
s = \theta_{1} - \theta_{\infty} = \theta_{\infty}e^{\frac{\bar{b} - 2\pi}{\bar{a}}},
\end{eqnarray}
\begin{eqnarray}
r_{m} = 2.5 \log \emph{R}_{0} = 2.5 \log (\frac{\mu_{1}}{\sum^{\infty}_{n = 2}\mu_{n}}) =\frac{ 5 \pi}{\bar{a}}\log e,
\end{eqnarray}
where $\emph{R}_{0}$ is a flux ratio between the first image and all the other images. By adopting all these observations such that $s$, $r_{m}$, and $\theta_{\infty}$, it is easy to evaluate  $\bar{a}$, $\bar{b}$ and $u_{ps}$ in the limit of strong deflection.
For the existence of that coupling in our universe, we compare the values of these observations to that of those observations which are predicted by the theoretical models of coupling. Due to this technique, it is easy to store the characteristics information in the strong gravitational lensing. However, the distance of our Galaxy is round about 8.5 kpc \cite{1} taken from the earth. So, the ratio becomes $GM/D_{1} \approx 2.4734 \times 10^{-11}$ and the Galactic central object mass is approximately evaluated to be $4.4 \times 10^{6} M_{\odot}$. The situation of photon coupling with Weyl tensor in a Kiselev spacetime shows that the values of the strong deflection limit functions $(\bar{a}, \bar{b})$ and other possible observables can easily be estimated numerically in strong gravitational lensing by solving the Eqs. (68), (93), (94) and (95). In Figs.8-10, we plot the dependenc of the observables $\theta_{\infty}$, $s$ and $r_{m}$ for different values of the Kiselev parameter $\sigma$ on the coupling constant $\alpha$. We observe that for the case of PPL the angular position of the observables $\theta_{\infty}$ and $r_{m}$ is the increasing function of coupling constant $\alpha$ but decreasing function of different values of $\sigma$. On the other hand in case of PPM the angular position of $\theta_{\infty}$ and $r_{m}$ is the directly decreasing function of both parameters $\alpha$ and $\sigma$, as shown in Fig.8 and Fig.10, respectively. The variation of angular separation $s$ is given in Fig.9, we see that for PPM case observable $s$ increases directly with $\alpha$ and $\sigma$, while in PPL this angular separation decreases with $\alpha$ and increases with $\sigma$, respectively.
\section{Conclusions}
This paper deals with the dynamical equation of photon coupled to Weyl tensor and the strong gravitational lensing in a Kiselev black hole spacetime. We find that the coupling parameter $\alpha$, Kiselev parameter $\sigma$ and the polarization directions are important for the advancement of coupled photons. These parameters also contribute significantly for explaining the photon sphere radius $r_{ps}$, angle of deflection, the coefficient $\bar{a}$ and $\bar{b}$ appearing in the lensing formula. The modified light cone conditions imply that in this spacetime, photons travel along null geodesics. Here, we conclude that when $\sigma$ tends to zero in Eq. (11), the critical value $\alpha_{c1} = M^{2}$ for PPM and  the critical value $\alpha_{c2} = \frac{-M^{2}}{2}$ for PPL, which is consistent with those in Schwarzschild black hole spacetime \cite{1}. From the equation of circular photon orbits, the radius $r_{ps}$ for PPM decreases
monotonously with the coupling parameter $\alpha$, while in the case of PPL, we find that $r_{ps}$ increases monotonously with $\alpha$. For PPM, the monotonicity of $r_{ps}$ gradually increases with $\sigma$ and for PPL, the monotonicity of $r_{ps}$ gradually decreases with $\sigma$ which is different from that in the Schwarzschild case \cite{1}.
The gravitational lensing formula functions $\bar{a}$ and $\bar{b}$ are given in Figs. 4 and 5, we see that with the increase of $\alpha$ the function $\bar{a}$ increases for PPM, while $\bar{a}$ decreases with $\alpha$ for PPL.
The function $\bar{b}$  for the case of PPM first decreases
down to its minimum with $\alpha$ and then increases up to its maximum with the further increase of $\alpha$; after that,
it decreases with $\alpha$ again. Meanwhile, for PPL the function $\bar{b}$ first increases
up to its maximum with $\alpha$ and then decreases down to its minimum with the further increase of $\alpha$; after that,
it increases with $\alpha$ again.
For PPM, the monotonicity of $\bar{a}$ directly increases respectively with $\sigma$ for PPM and PPL, which is different from that in the Schwarzschild case \cite{1}. Moreover, the variation of $\bar{b}$  with coupling parameter $\alpha$  for the different values of Kiselev parameter $\sigma$ for the case of PPL is totally converse to that for PPM.
The strong gravitational lensing $\alpha(\theta)$ have similar behaviors of the function $\bar{a}$. We obtain the shadow cast in a Kiselev black hole spacetime where the size of the shadow is a increasing function of both mass M and Kiselev parameter $\sigma$, respectively.
The variation of the angular separation $s$ with the coupling parameter $\alpha$ is converse to the variations of the relativistic images $\theta_{\infty}$ and the relative magnitude $r_{m}$ with the coupling parameter $\alpha$. The changes of these observables with $\alpha$ also depend on the value of
the Kiselev parameter $\sigma$. In the usual Kiselev black hole spacetime, the observables $\theta_{\infty}$ and the relative magnitude $r_{m}$ decreases with  $\sigma$, but $s$ increases, which is different from that in the Schwarzschild case \cite{1}.
\section*{Acknowledgments}
We are grateful to the scholarly anonymous referees, who put their efforts and give valuable suggestions for improving this manuscript.

\end{document}